\documentclass{aa501}
\usepackage{epsfig,times}

\makeatletter
\def\@cite#1#2{(#1\if@tempswa , #2\fi)}
\def\@citex[#1]#2{\if@filesw\immediate\write\@auxout{\string\citation{#2}}\fi
  \def\@citea{}\@cite{\@for\@citeb:=#2\do
    {\@citea\def\@citea{;\penalty\@m\ }\@ifundefined
       {b@\@citeb}{{\bf ?}\@warning
       {Citation `\@citeb' on page \thepage \space undefined}}%
\hbox{\csname b@\@citeb\endcsname}}}{#1}}
\makeatother

\title{X--ray and optical monitoring of the peculiar source\\
4U 1700+24/V934 Her\thanks{Partly based on observations collected
at the Bologna Astronomical Observatory in Loiano, Italy}}

\author{N. Masetti\inst{1}, D. Dal Fiume\inst{1}$^,$\thanks{Deceased},
G. Cusumano\inst{2}, L. Amati\inst{1}, C. Bartolini\inst{3}, S. Del
Sordo\inst{2}, F. Frontera\inst{1,4}, A. Guarnieri\inst{3}, M.
Orlandini\inst{1}, E. Palazzi\inst{1}, A.N. Parmar\inst{5}, A.
Piccioni\inst{3} and A. Santangelo\inst{2}
}

\institute{Istituto Tecnologie e Studio delle Radiazioni Extraterrestri,
CNR, via Gobetti 101, I-40129 Bologna, Italy
\and
Istituto di Fisica Cosmica ed Applicazioni all'Informatica, CNR, via ugo
La Malfa 153, I-90146 Palermo, Italy
\and
Dipartimento di Astronomia, Universit\`a di Bologna, via Ranzani 1,
I-40127 Bologna, Italy
\and
Dipartimento di Fisica, Universit\`a di Ferrara, via Paradiso 12, I-44100
Ferrara, Italy
\and
Astrophysics Division, Space Science Department of ESA, ESTEC, Postbus
299, NL-2200 AG Noordwijk, The Netherlands
}

\offprints{Nicola Masetti, masetti@tesre.bo.cnr.it}
\date{Received on September 19, 2001; Accepted on November 7, 2001}

\begin{document}

\abstract{
We report on {\it ASCA} and {\it BeppoSAX} X--ray broad band observations
of the galactic low-luminosity X--ray source 4U 1700+24 performed on 1995
and 1998, respectively, and on (quasi-)simultaneous ground
observations of its optical counterpart, V934 Her, from the
Loiano 1.5-meter telescope. In order to better understand the nature of 
the source we also analyze public archival {\it ROSAT} and {\it RXTE} 
data as well as the {\it RXTE} ASM light curve of 4U 1700+24; we also 
re--analyze a 1985 {\it EXOSAT} pointing. The optical spectra are typical 
of a M2 III star; this allows us to determine a revised distance to the
object of $\sim$400 pc. While these spectra do not show either any
spectral change between the two epochs or any peculiar feature apart from
those observed in normal red giants, the spectroscopic measurements
carried out in X--rays reveal a complex and long-term variable spectrum,
with a clear soft excess. The X--ray spectral properties of the source 
are best described by a thermal Comptonization spectrum plus a soft energy 
($<$1 keV) excess, which can be modeled in the form of a blackbody emission
with $kT_{\rm BB} \sim$ 1 keV; the latter component is not detected at
the lowest source flux levels. The ratio between the two components varies
substantially with the source flux. The X--ray emission from the 
object appears to become harder as its luminosity increases:
indeed, the {\it RXTE} data acquired during an outburst occurred in
October-November 1997 display a hard tail, detected up to 100 keV and
modeled with a comptonizing cloud which is hotter and less opaque than that
seen in the low intensity state.
Apart from erratic shot-noise variability on timescales of tens to thousands 
of seconds, no significant properties (such as pulsations or QPOs) are
found from the timing analysis of the X--ray light curves extracted from
the observations presented here.
With the new distance determination, the 2--10 keV X--ray luminosity range
spanned in the considered observations lies between
$\sim$2$\times$10$^{32}$ and $\sim$1$\times$10$^{34}$ erg s$^{-1}$.
All this information, combined with the findings by other authors, allows
us to suggest that the scenario which best describes the object consists
of a wide binary system in which a neutron star accretes matter from the
wind of a M-type giant star. Implications of such a model are discussed.
}

\maketitle
\markboth{N. Masetti et al.: X--ray and optical monitoring of 
4U 1700+24}{N. Masetti et al.: X--ray and optical monitoring of 
4U 1700+24}

\section{Introduction}

In optical astronomy the identification of a binary system derives in most
cases from the observation of photometric and/or radial velocity
variations at the orbital period; these techniques also allow determining
the masses of the two components. This application is of great importance
in X--ray binaries, as it helps understanding the nature of the accreting
compact object (hereafter labeled also as `primary component', in contrast 
with the mass donor star which is indicated as `secondary component'). 
As not all X--ray binaries have optical counterparts known or displaying
periodic modulation associated with the orbital period, a further
effective criterium in galactic X--ray astronomy for the identification of
a binary system harbouring an accreting compact object is often based on
the observed X--ray luminosity. For persistent X--ray binaries with
a neutron star (NS) or possibly a black hole (BH), luminosities $L_{\rm
X}$ of the order of at least 10$^{34}$ -- 10$^{35}$ erg s$^{-1}$ are
easily reached.
The diagnosis of the presence of a NS rather than a BH in bright
persistent Low Mass X--Ray Binaries (LMXRBs) is in most cases driven by
the detection of pulsations or thermonuclear X--ray bursts. 
X--ray binaries harbouring white dwarfs (WDs) also show some distinctive
features. As an example, in polars and intermediate polars optical and
UV observations often reveal the typical signatures of the presence of a
WD in the system. 
Orbital periods and light curves also add unambiguous and reliable
evidence of the presence of WDs in this class of X--ray binaries.

For a number of X--ray sources (see e.g. the catalog of Liu et al. 2001)
the identification of a class or even the
diagnosis of binarity is rather difficult, especially when the observed
X--ray luminosity is $L_{\rm X}$ $\leq 10^{33}$ erg s$^{-1}$, as they 
do not easily fit in any of the groups of objects just listed.
They do not show any signature in the optical radial velocity to 
allow the determination of an orbital period, and obviously of the system 
mass function. Also, they often do not seem to have pulsations or modulations 
in their X--ray flux which could be related to rotational or orbital periods.
They are associated with Population I stars, therefore suggesting a  
relatively young age, and show persistent emission with a X--ray luminosity 
varying by at least an order of magnitude. They also display irregular
short-term variability, from thousands to tens of seconds and sometimes
shorter. The object 4U 1700+24 belongs to this group.

This X--ray source was first detected as a variable object by the
satellites {\it Ariel V} (Cooke et al. 1978) and {\it Uhuru}
(Forman et al. 1978). It shows extremely erratic variations, but no
pulsations were ever detected. The rapid (10--1000 s) time variability in
X--rays observed with {\it Einstein} and {\it EXOSAT} spacecraft is
strongly suggestive of turbulent accretion, often observed in X--ray
binaries; moreover, the X--ray spectrum of 4U 1700+24 appears rather
energetic and was measured up to 10 keV (Garcia et al. 1983; Dal Fiume et
al. 1990). The hardness of the spectrum suggests an origin from
accretion onto a compact object, but it cannot be used as a decisive
parameter. 

Its optical counterpart, the bright ($V \sim$ 7.8) red star
HD154791 (subsequently named V934 Her; Kazarovets
et al. 1999), was identified by Garcia et al. (1983) as a late type giant 
of spectral class M3 II on the basis of the positional coincidence
with the error boxes determined by the {\it Einstein} and {\it HEAO1}
satellites. The optical spectrum of this giant looks quite typical
(Garcia et al. 1983; Dal Fiume et al. 1990), with no signs of
peculiarities such as emission lines; this fact makes
this system unique in the sense that this is the only LMXRB with a
``normal'' giant star as optical counterpart. However, Gaudenzi \&
Polcaro (1999) recently found some interesting and variable features in
its optical spectrum; these authors gave also a different spectral
classification
of this star (M3 III). Variable UV line emission was detected by Garcia et
al. (1983) and by Dal Fiume et al. (1990) in different IUE pointings,
showing at last some unusual features in the emission from this otherwise
normal giant. This energetic emission is likely linked to the same
mechanism that produces the observed X--ray emission.
The X--ray luminosity of the system, computed assuming a distance of 730
pc (Garcia et al. 1983), is $L_{\rm X}$ $\sim$ 10$^{33}$ erg s$^{-1}$.
Infrared emission from this object was observed up to 25 $\mu$m
with {\it IRAS} (Schaefer 1986), while no counterpart was detected
at radio wavelengths (Wendker 1995).

Recently, during the Fall of 1997, an X--ray outburst peaking at
$\approx$30 mCrab in the 2--10 keV band was detected with
{\it RXTE} from 4U 1700+24 (see the long--term X--ray light curve reported 
in Fig. 1); optical spectra acquired during outburst did not show
any change with respect to those obtained during low intensity state 
(Tomasella et al. 1997).
Finally, in spite of various attempts, no evidence of an orbital period
was obtained from radial velocity analysis of optical spectra (e.g. Garcia
et al. 1983).
Therefore, the picture emerging from observations gives only hints in  
favour of a binary system, as no ``classical'' features to be
associated with the presence of a compact object were ever found.

The aim of this paper is therefore to get a deeper insight on the
high-energy properties of this source and to test whether the binary M
giant plus accreting compact object scenario for this system is viable. 
To this task we have monitored this source over a time span of $\sim$15
years with various X--ray satellites ({\it EXOSAT}, {\it ASCA} and {\it
BeppoSAX}). Additionally, in order to achieve a better understanding of
the long- and short-term X--ray behaviours of this source, we analyzed
still unpublished archival {\it ROSAT} and {\it RXTE} public data
and we retrieved the data collected with the {\it RXTE} All-Sky Monitor
(ASM). Also, here we report on ground-based optical spectroscopic 
observations taken at the Loiano 1.5-meter telescope of the Bologna
Astronomical Observatory (quasi-)simultaneously with the {\it ASCA} and
{\it BeppoSAX} pointings.

The paper is organized as follows: Sect. 2 will illustrate the
observations and the data analysis, while in Sect. 3 the X--ray and 
optical results will be reported; in Sect. 4 a discussion will be given,
and in Sect. 5 we will draw the conclusions.

\begin{table*}
\caption[]{Journal of the X--ray pointings on 4U 1700+24 analyzed in
this paper}
\begin{center}
\begin{tabular}{llcrrl}
\noalign{\smallskip}
\hline
\noalign{\smallskip}
\multicolumn{1}{c}{Satellite} & \multicolumn{1}{c}{Obs. start} &
Obs. start & \multicolumn{1}{c}{Exposure} & 
\multicolumn{2}{l}{On-source time (ks)} \\
 & \multicolumn{1}{c}{date} & time (UT) & \multicolumn{1}{c}{(ks)} & 
& \\
\noalign{\smallskip}
\hline
\noalign{\smallskip}
{\it EXOSAT}   & 1985 Mar 15 & 00:18:24 &   40.1 & 36.3 & (ME)$^*$\\
\noalign{\smallskip}
{\it ROSAT}    & 1992 Mar 22 & 23:07:33 &  105.0 &  7.9 & (PSPC B)\\  
               & 1997 Sep 8  & 13:07:32 & 1164.8 & 10.6 & (HRI)\\
\noalign{\smallskip}
{\it ASCA}     & 1995 Mar 8  & 11:39:26 &   41.5 & 22.5 & (GIS2) \\
	       &             & 11:39:26 &   41.5 & 22.5 & (GIS3) \\
	       &             & 11:39:16 &   41.5 & 21.8 & (SIS0) \\
 	       &             & 11:39:16 &   41.5 & 21.8 & (SIS1) \\
\noalign{\smallskip}
{\it RXTE}     & 1997 Feb 14 & 06:56:32 &   25.3 & 11.5 & (PCA)\\
               & 1997 Nov 15 & 11:40:00 &   12.9 &  7.9 & (PCA)\\
               & 1997 Nov 15 & 11:40:00 &   12.9 &  2.5 & (HEXTE)\\
\noalign{\smallskip}
{\it BeppoSAX} & 1998 Mar 27 & 13:47:22 &   46.0 & 12.1 & (LECS)\\
	       &	     & 13:47:22 &   46.0 & 24.9 & (MECS)\\
 	       &	     & 13:47:22 &   46.0 & 12.4 & (HPGSPC)\\
	       &             & 13:47:22 &   46.0 & 11.8 & (PDS)\\
\noalign{\smallskip}
\hline
\noalign{\smallskip}
\multicolumn{6}{l}{$^*$Results from this observation were already
presented by Dal Fiume et al. (1990)}\\
\noalign{\smallskip}
\hline
\noalign{\smallskip}
\end{tabular}
\end{center}
\end{table*}

\section{Observations and data reduction}

\subsection{{\it ASCA} and {\it BeppoSAX} data}

In this Subsection and in the following one we describe the X--ray
observations made with several spacecraft on 4U 1700+24. 
Table~1 reports the log of all the X--ray observations
presented in Sections 2.1 and 2.2.

4U 1700+24 was observed with the satellite {\it ASCA} (Tanaka et al. 1994)
on 1995 March 8.
This spacecraft carried four X--ray telescopes (Serlemitsos et al. 1995)
equipped with two Gas Imaging Spectrometers (GIS; Ohashi et al.
1996, Makishima et al. 1996) and two Solid-State Imaging Spectrometers
(SIS; Yamashita et al. 1997), each of them sensitive in the 0.5--10 keV
range. 
Standard filtering criteria were applied to the SIS and GIS data:
we selected time intervals outside the South Atlantic Geomagnetic Anomaly
(SAGA) and for which the elevation angle above the earth
limb was $>$$5^{\circ}$ and the cutoff rigidity was below 7 GV.
Data were extracted from a circular area centered on the source position
and with radius 6$'$ for the GIS and 3$'$ for the SIS.
The background was evaluated from blank areas of GIS and SIS images
of the 4U 1700+24 field and then subtracted from the source events.

This source was also observed with the Narrow Field Instruments (NFIs)
onboard {\it BeppoSAX} (Boella et al. 1997a) on 1998 March 27.
The NFIs include the Low-Energy Concentrator Spectrometer (LECS,
0.1--10~keV; Parmar et al. 1997), two Medium-Energy Concentrator
Spectrometers (MECS, 1.5--10~keV; Boella et al. 1997b), a High Pressure
Gas Scintillation Proportional Counter (HPGSPC, 4--120~keV; Manzo et al.
1997), and the Phoswich Detection System (PDS, 15--300~keV; Frontera et
al. 1997). 
During all pointings the four NFIs worked nominally and the source was
detected by all of them. 
Good NFI data were selected from intervals outside the SAGA
when the elevation angle above the earth limb was $>$$5^{\circ}$, when
the instrument functioning was nominal and, for LECS events, during
spacecraft night time. The SAXDAS 2.0.0 data
analysis package (Lammers 1997) was used for the extraction and
the processing of LECS, MECS and HPGSPC data.
The PDS data reduction was instead performed using XAS version 2.1
(Chiappetti \& Dal Fiume 1997).
LECS and MECS data were reduced using an extraction radius of 4$'$
and 3$'$, respectively, centered at the source position; before
extraction, data from the two MECS units were merged together.
Background subtraction for the two imaging instruments was performed using
standard library files, while the background for the HPGSPC and for the
PDS data was evaluated from the fields observed during off-source pointing
intervals.

\begin{table}
\caption[]{Journal of the optical spectroscopic observations of V934 Her
presented in this paper}
\begin{center}
\begin{tabular}{lcccc}
\noalign{\smallskip}
\hline
\noalign{\smallskip}
\multicolumn{1}{c}{Date} & Obs. start & Exposure & Passband & Slit \\
 & time (UT) & time (s) & (\AA) & width ($''$) \\
\noalign{\smallskip}
\hline
\noalign{\smallskip}

 1995 Mar 8  & 01:09:03 & 120 & 4000--8000 & 1.5 \\
 1998 Mar 28 & 03:37:14 & 300 & 4000--8000 & 2.0 \\
 1998 Mar 28 & 03:43:56 & 300 & 4000--8000 & 2.0 \\

\noalign{\smallskip}
\hline
\noalign{\smallskip}
\end{tabular}
\end{center}
\end{table}

\subsection{Other public X--ray data}

In order to present a thorough long-term analysis of the X--ray
behaviour of 4U 1700+24, we also retrieved from the HEASARC
archive\footnote{available at:~{\tt http://heasarc.gsfc.nasa.gov/\\
cgi-bin/W3Browse/w3browse.pl}}
all the unpublished X--ray observations on this source. These comprise
{\it ROSAT} and {\it RXTE} data; we also re-examine 
an {\it EXOSAT} pointing already presented in Dal Fiume et al. (1990).

Two pointed observations were performed on 4U 1700+24 with the {\it ROSAT}
satellite (Tr\"umper 1982). The first one was acquired between March 22
and 24, 1992 with the Position Sensitive Proportional Counter (PSPC;
Pfeffermann \& Briel 1986) unit B, sensitive in the 0.1--2.4 keV range.
The second observation was performed with the High Resolution Imager (HRI;
Pfeffermann et al. 1987) onboard {\it ROSAT} between September 8 and
22, 1997, again in the 0.1--2.4 keV range. The HRI detector had a better
spatial resolution than PSPC, but it had no spectral resolution, so it
could  not provide any spectral information on the source.
PSPC spectra and light curves were selected by extracting events from a
region centered on the source centroid and with 1\farcm8 radius; good HRI
data were instead selected from a circular region with 45$''$ radius
centered on the source. The background was evaluated from blank
selected areas in the observed field.

The {\it RXTE} satellite (Bradt et al. 1993) carries a 5-unit Proportional
Counter Array (PCA; Jahoda et al. 1996), which is sensitive in the 2--60
keV energy range and allows a time resolution of 1 $\mu$s, and a High
Energy X--ray Timing Experiment (HEXTE; Rothschild et al. 1998) composed
of two clusters of 4 phoswich scintillation detectors working in the
15--250 keV energy range. This satellite also carries an
ASM\footnote{ASM light curves are available at:\\ 
{\tt http://xte.mit.edu/ASM\_lc.html}} (Levine et al. 1996) which regularly
scans the X--ray sky in the 2--10 keV range with a daily sensitivity
of 5 mCrab. Figure 1 reports the complete up-to-date (October 2001) 
2--10 keV ASM
light curve of 4U 1700+24 starting on January 1996, along with the times
of the single {\it BeppoSAX} pointing, of the second {\it ROSAT} one and
of the two {\it RXTE} pointed observations. In order to clearly display the
long-term trend in the X--ray emission from 4U 1700+24, each point of Fig.
1 is computed as the average of 5 subsequent measurements; given that the
single original ASM points we retrieved were basically acquired on a daily
basis (i.e. they are one-day averaged measurements), the plot illustrated 
in Fig. 1 roughly corresponds to a 5-day averaged X--ray light curve.

\begin{figure*}
\vspace{-4cm}
\begin{center}
\epsfig{figure=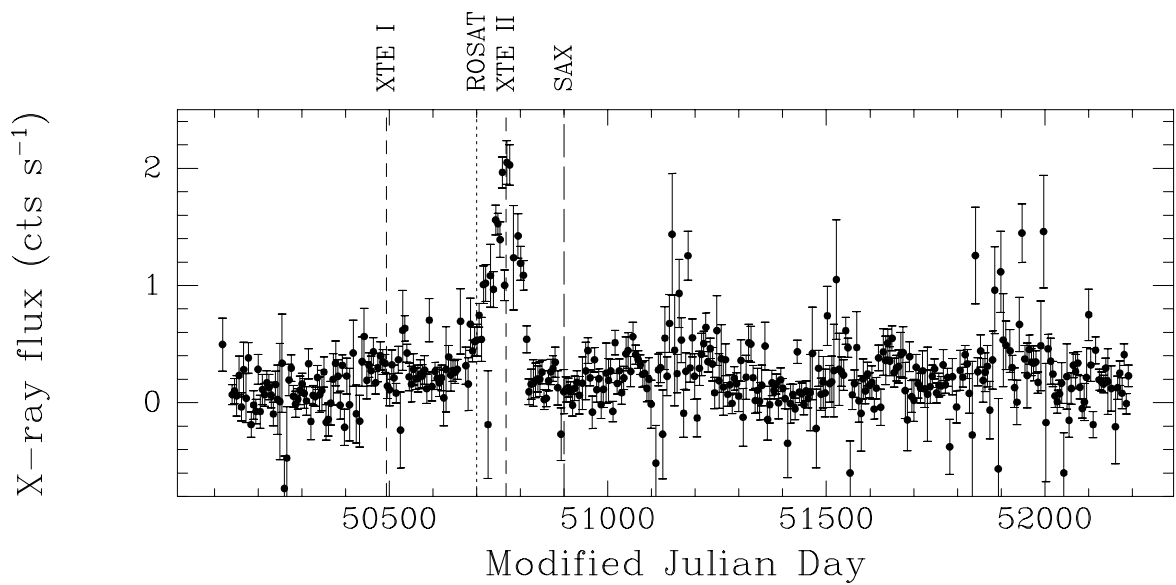,width=19cm,angle=270}
\end{center}
\vspace{-7cm}
\caption[]{2--10 keV 5-day averaged {\it RXTE} ASM light curve of
4U 1700+24. 1 ASM count s$^{-1}$ roughly corresponds to 13 mCrab assuming
a Crab-like spectrum. In the plot the times of the pointed {\it RXTE},
{\it ROSAT} and {\it BeppoSAX} are indicated by the vertical dashed lines
(different dashes correspond to different spacecraft). After the bigger
outburst occurred around MJD 50750, a series of lower intensity
quasiperiodic increases in the X--ray activity from the source are
noticed with a recurrence time of about 400 days}
\end{figure*}

The {\it RXTE} pointings were performed on 4U 1700+24 on 1997 February 
14 and on 1997 November 15; the latter observation was acquired during the
fairly intense X--ray outburst detected by the ASM (see Fig. 1). In the
following we will refer to these two {\it RXTE} pointings as observations
I and II, respectively.
The PCA instrument detected 4U 1700+24 in both occasions; during the two
observations, all five detector units of the PCA were on. Good PCA data
were then selected after filtering for low-earth events and high
background times. The background was subtracted using the standard PCA
background models. Instead, HEXTE detected the source during the second
pointing only, when the source was in outburst. In this case, a similar
procedure as above was applied to each HEXTE cluster to select good data.
Background was determined by periodically rocking the cluster elements on-
and off-source.
The spectral analysis was performed on good data extracted in 
the {\it RXTE}~``Standard 2" mode, which provides full spectral
information with a time resolution of 16 s.

Finally, the {\it EXOSAT} observation was acquired on 1985 March 15 
with the Medium Energy (ME; Turner et al. 1981) proportional counter 
(1--50 keV). The source was detected only up to 10 keV during this 
observation. A more detailed description on this observation and on the 
applied reduction techniques can be found in Dal Fiume et al. (1990).

\subsection{Optical data}

Optical medium-resolution spectra of the star V934 Her were acquired,
(quasi-)simultaneously with the {\it ASCA} and {\it BeppoSAX} pointings,
in Loiano (Italy) with the Bologna Astronomical Observatory 1.52-meter
``G.D. Cassini" telescope plus BFOSC on 1995 March 8 and on 1998 March 28.
The telescope was equipped with a 1100$\times$1100 pixels Thompson CCD on
both occasions. Grism \#4 was used on both epochs, providing the
nominal spectral coverage indicated in Table 2. The slit width was
1$\farcs$5 for the March 1995 observation and 2$''$ for the March 1998
pointing. The use of this setup secured a final dispersion of 3.7 \AA/pix
for the spectra acquired during both nights.
The complete log of the optical spectroscopic observations presented 
here is reported in Table 2.

Spectra, after correction for flat-field and bias, were background
subtracted and optimally extracted (Horne 1986)
using IRAF\footnote{IRAF is the Image
Analysis and Reduction Facility made available to the astronomical
community by the National Optical Astronomy Observatories, which are
operated by AURA, Inc., under contract with the U.S. National Science
Foundation. STSDAS is distributed by the Space Telescope Science
Institute, which is operated by the Association of Universities for
Research in Astronomy (AURA), Inc., under NASA contract NAS 5--26555.}.
Helium-Argon lamps were used for wavelength calibration; spectra taken 
on March 1998 were then flux-calibrated by using the spectrophotometric
standard HD161817 (Philip \& Hayes 1983; Silva \& Cornell 1992) and
finally averaged together. Correction for slit losses was also introduced.
For the March 1995 spectrum no spectroscopic standard was available. 
The correctness and consistency of March 1998 spectra flux calibration was 
checked against the optical photoelectric data collected in the $UBVRI$ bands
with the 0.6-m Loiano telescope at the epoch in which these spectra were 
acquired (Diodato 1998).
Wavelength calibration was instead checked by using the positions of 
background night sky lines: the error was 0.5 \AA.

\begin{figure*}
\vspace{-5cm}
\begin{center}
\epsfig{figure=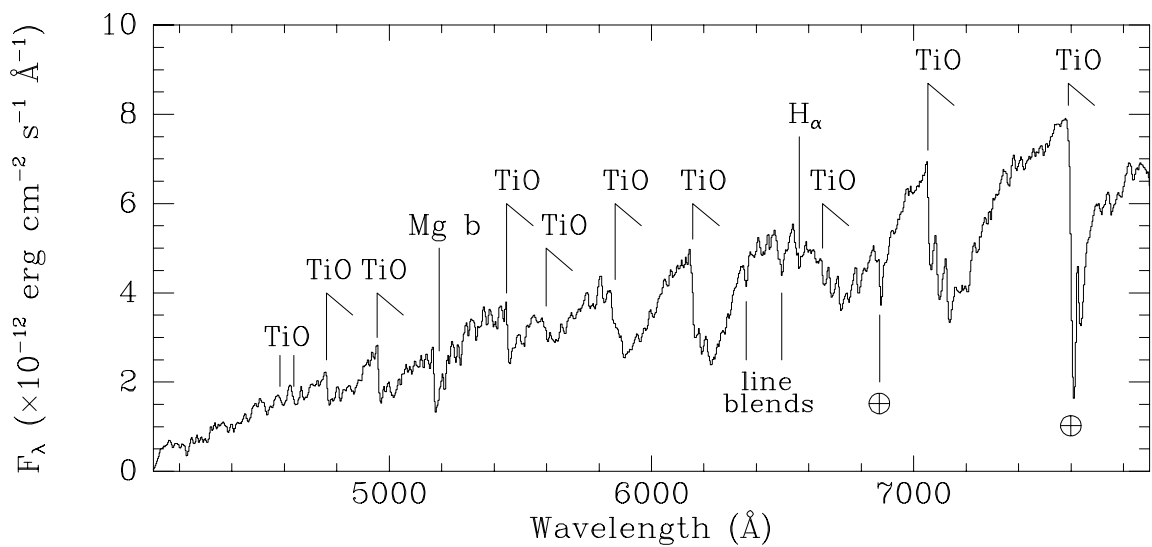,width=19cm,angle=270}
\end{center}
\vspace{-7cm}
\caption[]{4100-7900 \AA~optical spectum of V934 Her obtained 
with the Loiano 1.52-meter telescope plus BFOSC on 1998 March 28 and
dereddened using $E(B-V)$ = 0.044. The spectrum is typical of
a star of type M2 III (see text). 
The telluric absorption bands at 6870 and 7600 \AA~are marked with the
symbol $\oplus$.}
\end{figure*}

\section{Results}

\subsection{Optical spectra and a new distance determination}

The flux-calibrated averaged 4100--7900 \AA~spectrum of V934 Her acquired
on 1998 March 28 is reported in Fig. 2. As it appears evident, it is
dominated by TiO absorption bands and no emission features are apparent.
This is typical of normal M-type giants, as already remarked by other
authors (Garcia et al. 1983; Gaudenzi \& Polcaro 1999).
The H$_\alpha$ line is also detected in absorption.
We do not detect the emission wings reported by Gaudenzi \& Polcaro (1999)
for this line. This might suggest that these wings are variable in
intensity due to e.g. changes in the X--ray irradiation (see below), but 
their non-detection in our spectra might however be also due to our 
coarser spectral dispersion (3.7 \AA/pix, against 1.5 \AA/pix
of the 6300-7800 \AA~spectra of Gaudenzi \& Polcaro 1999).
We also detect, among the main spectral features, the Mg `b' absorption
around 5170 \AA~and two atomic line blends of metal intersystem lines
of Fe {\sc i}, Ti {\sc i}, Cr {\sc i}, Ba {\sc i}, Ca {\sc i}, Mn {\sc i},
Co {\sc i} and Ni {\sc i} located at 6352 \AA~and 6497 \AA~(see e.g. 
Turnshek et al. 1985). Telluric absorption features are moreover detected
at 6870 and 7600 \AA.

We next compare the March 1998 averaged spectrum with the one taken on
March 1995. As the latter could not be calibrated in flux, we
normalized both to their continuum. No appreciable variation in the
strength of any absorption line or band between the two epochs is
detected in the considered spectral range.

Then, using the Bruzual-Persson-Gunn-Stryker Spectrophotometry
Atlas\footnote{available at:\\
{\tt ftp://ftp.stsci.edu/cdbs/cdbs1/grid/bpgs/}}
(Gunn \& Stryker 1983), we compare the spectrum of V934
Her with that of several red stars whose spectra are present in this
atlas. To this task we dereddened our spectrum assuming $E(B-V)$ =
0.044; this value is derived from the dust maps by Schlegel et al. (1998).
The best match is obtained with the M2 III star HD104216, whose
optical spectrum is strikingly similar to that of V934 Her. A good match,
albeit somewhat poorer, is obtained with the star HD142804 (M1 III).
Stars of later or earlier spectral types, as
well as (see the classification by Garcia et al. 1983) an M3 II template,
the star BD +19$^{\circ}$1947 from the Jacoby-Hunter-Christian
Spectrophotometric Atlas\footnote{available at:\\ {\tt
ftp://ftp.stsci.edu/cdbs/cdbs1/grid/jacobi/}} (Jacoby et al.
1984), provide substantially poorer matches.
Thus, we can confidently state that the spectral type of V934 Her is M2
III. This is consistent with the classification
made by Gaudenzi \& Polcaro (1999), albeit ours appears slightly bluer.

Concerning the more significant difference between our classification and
that provided by Garcia et al. (1983), it could be suggested that the
star underwent a consistent spectral change somewhere in the past between
the years of Garcia et al.'s (1983) and of
Gaudenzi \& Polcaro's (1999) spectral observations. We however consider
this explanation unlikely on the basis of the lack of a strong optical 
variability and color changes in the $UBVRI$ photometry (Dal Fiume et al.
2000; Diodato 1998), which always shows colors
typical of a luminosity class III M-type giant (Lang 1992).
Thus, we believe that the spectral classification of V934 Her made by
Garcia et al. (1983) using the width of the Ca {\sc ii} K emission line 
could be biased by the
UV variability of this object and by the possible presence of the
additional absorption suggested by Gaudenzi \& Polcaro (1999). Also,
from the data set of Engvold \& Rygh (1978), using which
Garcia et al. (1983) determined the absolute magnitude M$_V$ in the $V$
band (and then the spectral type) of V934 Her, it seems that the relation
between this quantity and the width of the Ca {\sc ii} K line is very
steep for giant stars, i.e. a small variation in the line width reflects
in a substantial jump of M$_V$. Therefore we consider the comparison with
template optical spectra of stars as a more secure method of spectral
classification.

Assuming that V934 Her is a star of spectral type M2 III, we can revise
its distance in the following way. Given that the absolute magnitude M$_V$
of a M2 III star is $-$0.6 (Lang 1992), and that the $V$ magnitude
of V934 Her, corrected for interstellar absorption, is $V\sim$ 7.5, we 
obtain, by applying the distance modulus formula, $d$ = 420$\pm$40 pc;
here the error reflects the uncertainties on the intervening galactic 
absorption, on the mean $V$-band magnitude of the star and on the
spectral classification.
This value is fully consistent with the distance $d$ = 390$\pm$130 pc
obtained from the annual parallax ($\pi$ = 2.57$\pm$0.86 milliarcsec)
of this source as measured with the {\it Hipparcos} satellite (Perryman
et al. 1997).
In our distance estimate we made the assumption that all of the $V$-band
magnitude comes from the optical companion. This, given the shape of the
optical spectrum, is in our opinion reasonable.

Thus, throughout the rest of the paper we will use our revised distance of
420 pc for the 4U 1700+24 X--ray luminosity measurements.

\begin{figure*}
\begin{center}
\epsfig{figure=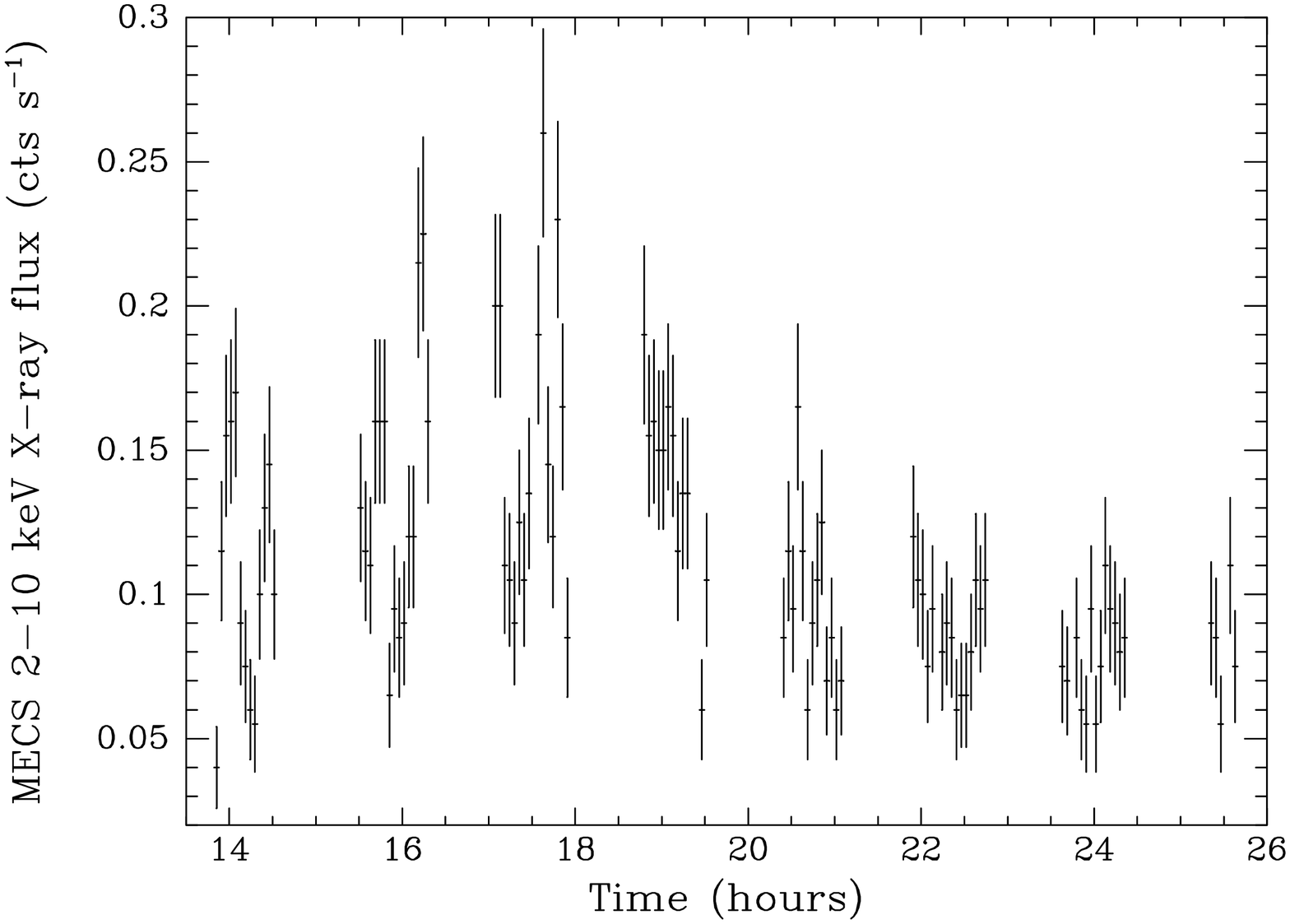,width=9.8cm,angle=-90}
\hspace{1.5cm}
\epsfig{figure=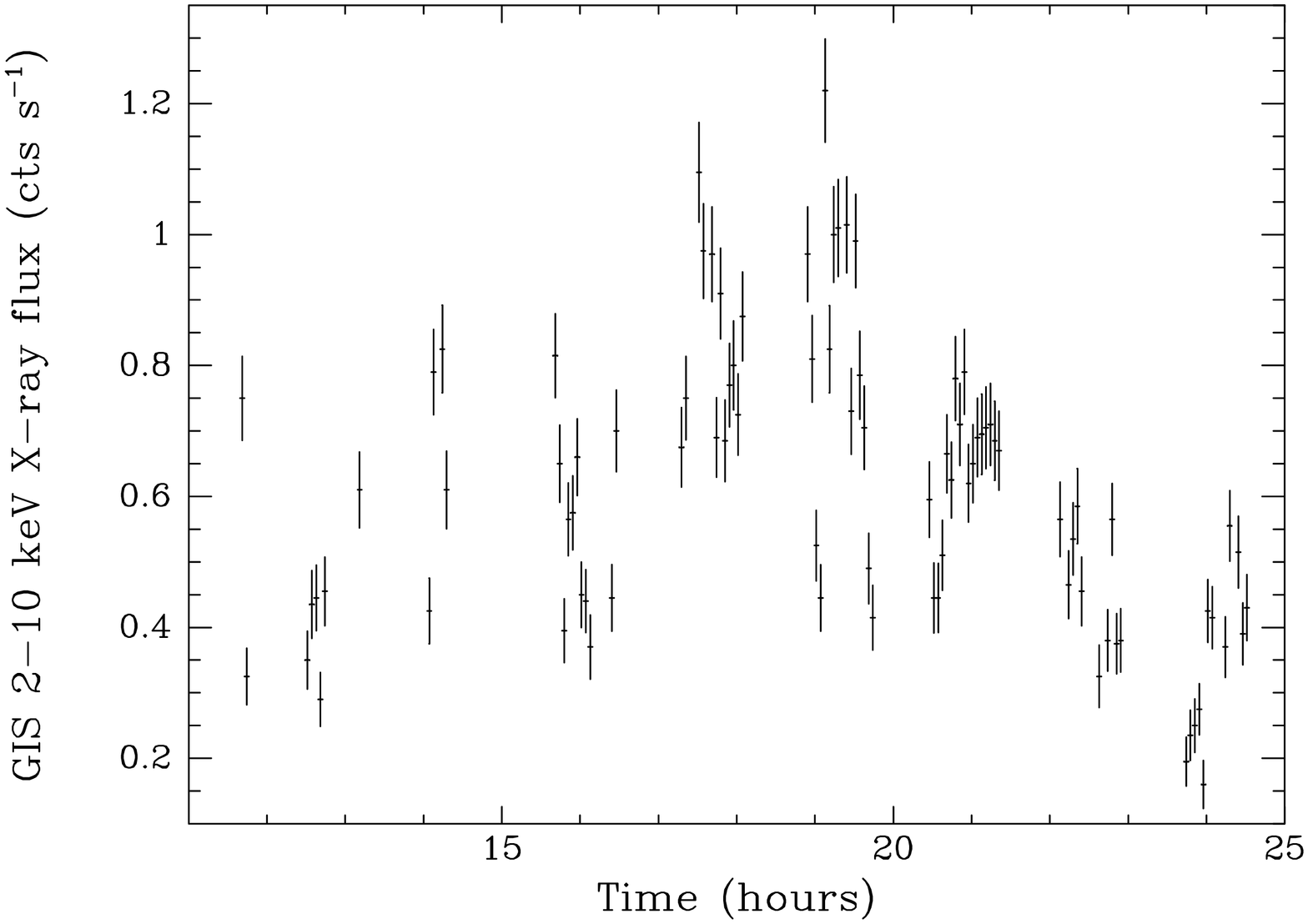,width=9.8cm,angle=-90}
\end{center}
\vspace{-4.5cm}
\begin{center}
\epsfig{figure=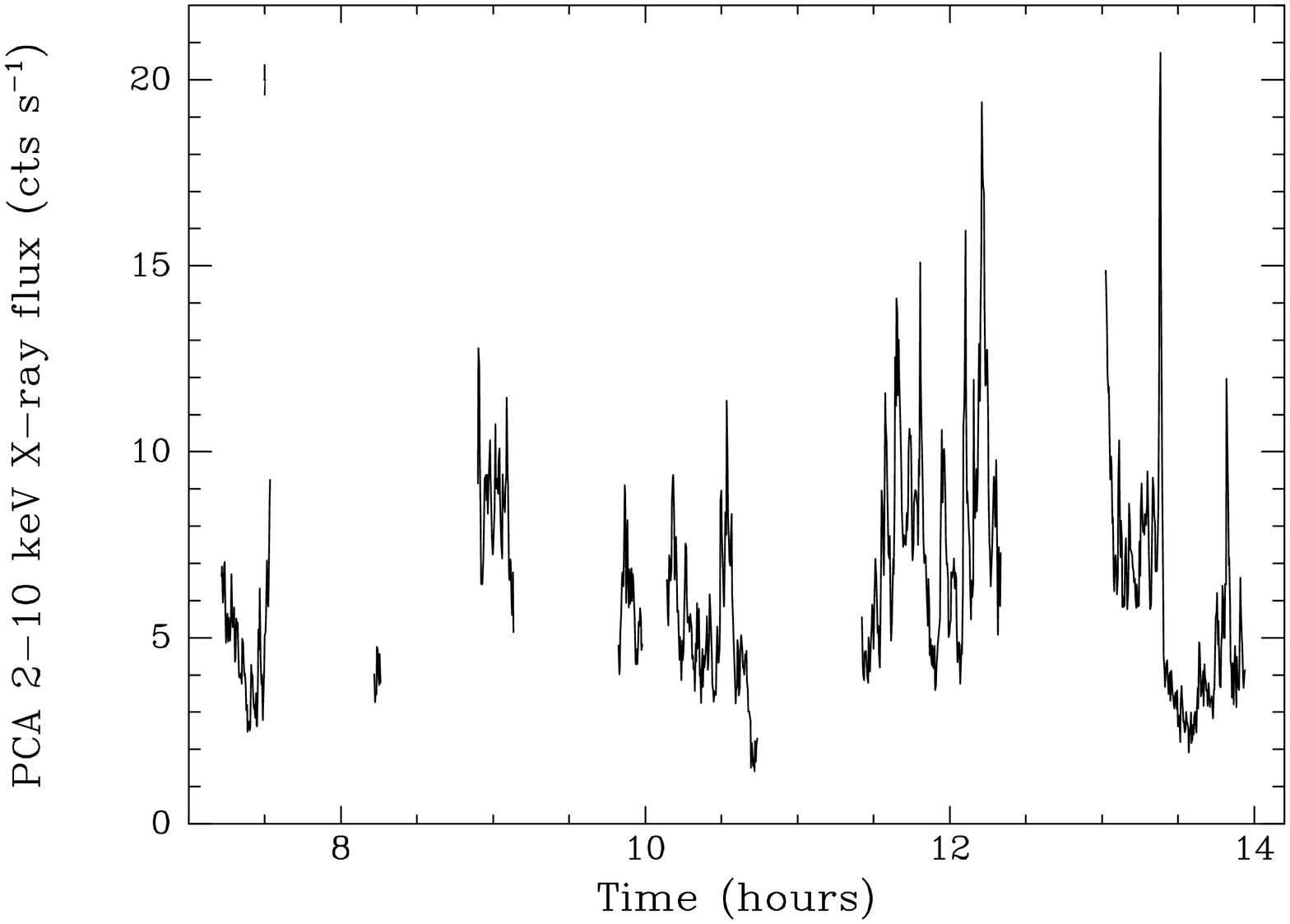,width=9.8cm,angle=-90}
\hspace{1.5cm}
\epsfig{figure=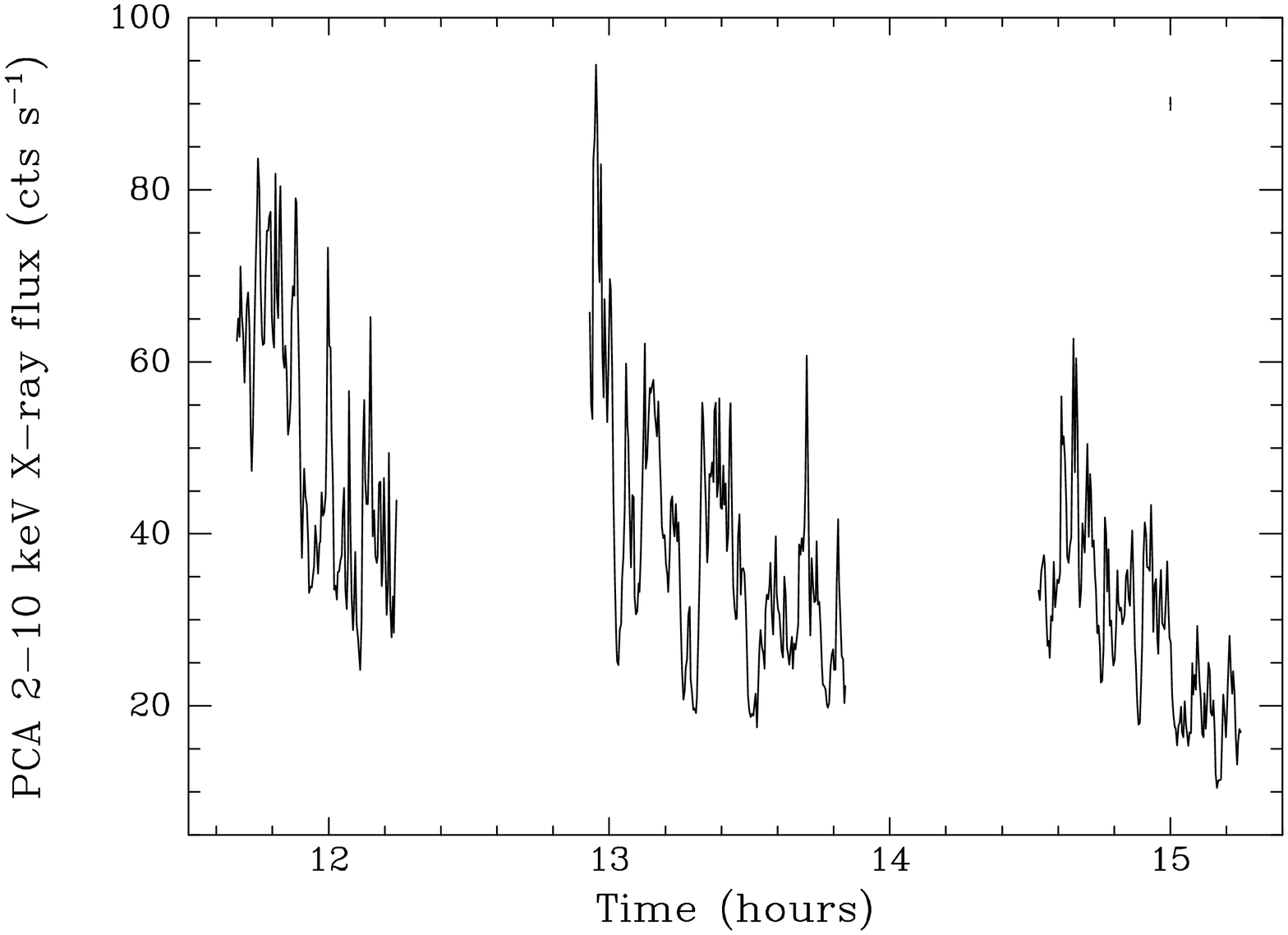,width=9.8cm,angle=-90}
\end{center}
\vspace{-4.5cm}
\begin{center}
\epsfig{figure=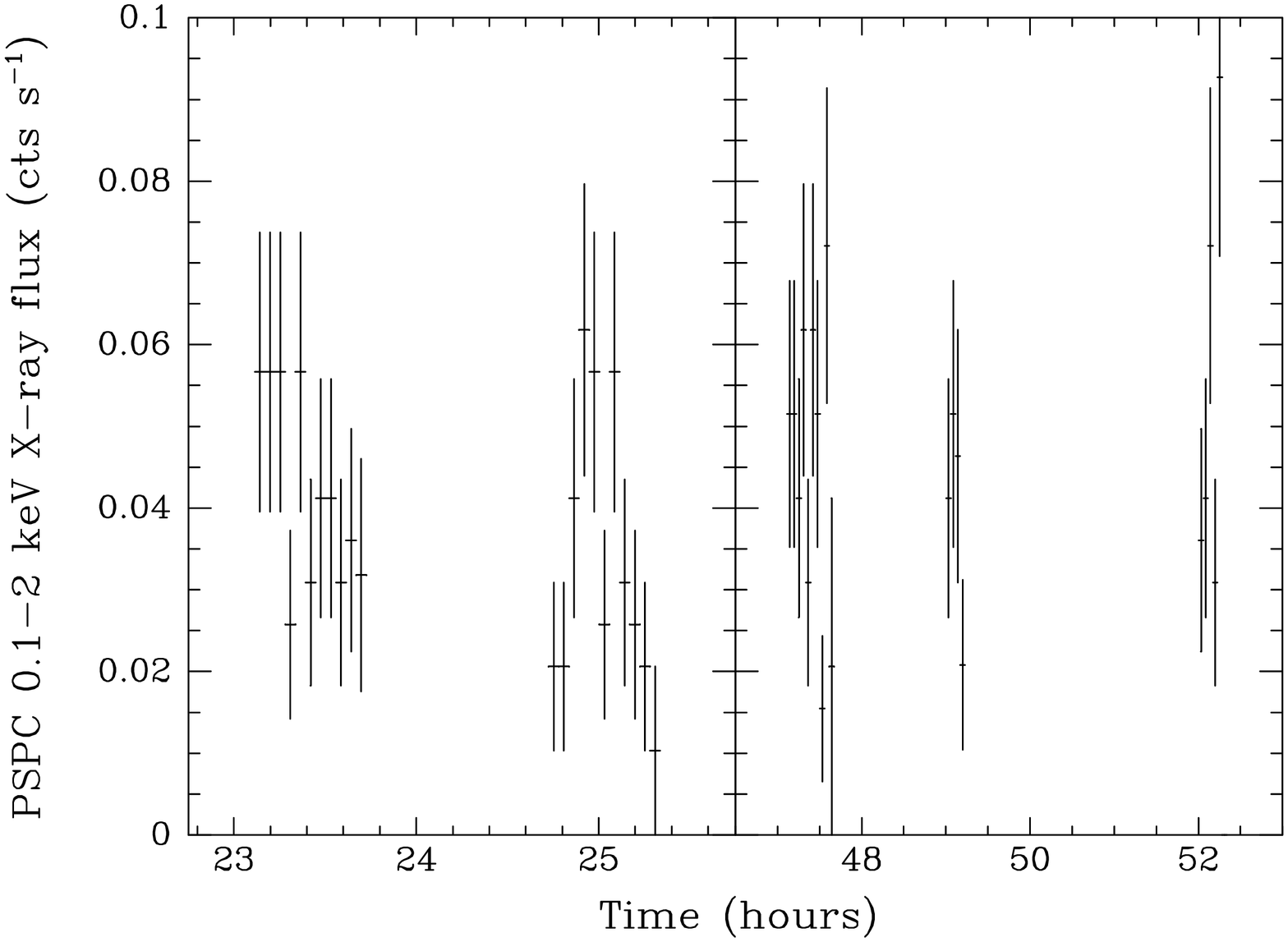,width=9.8cm,angle=-90}
\hspace{1.5cm}
\epsfig{figure=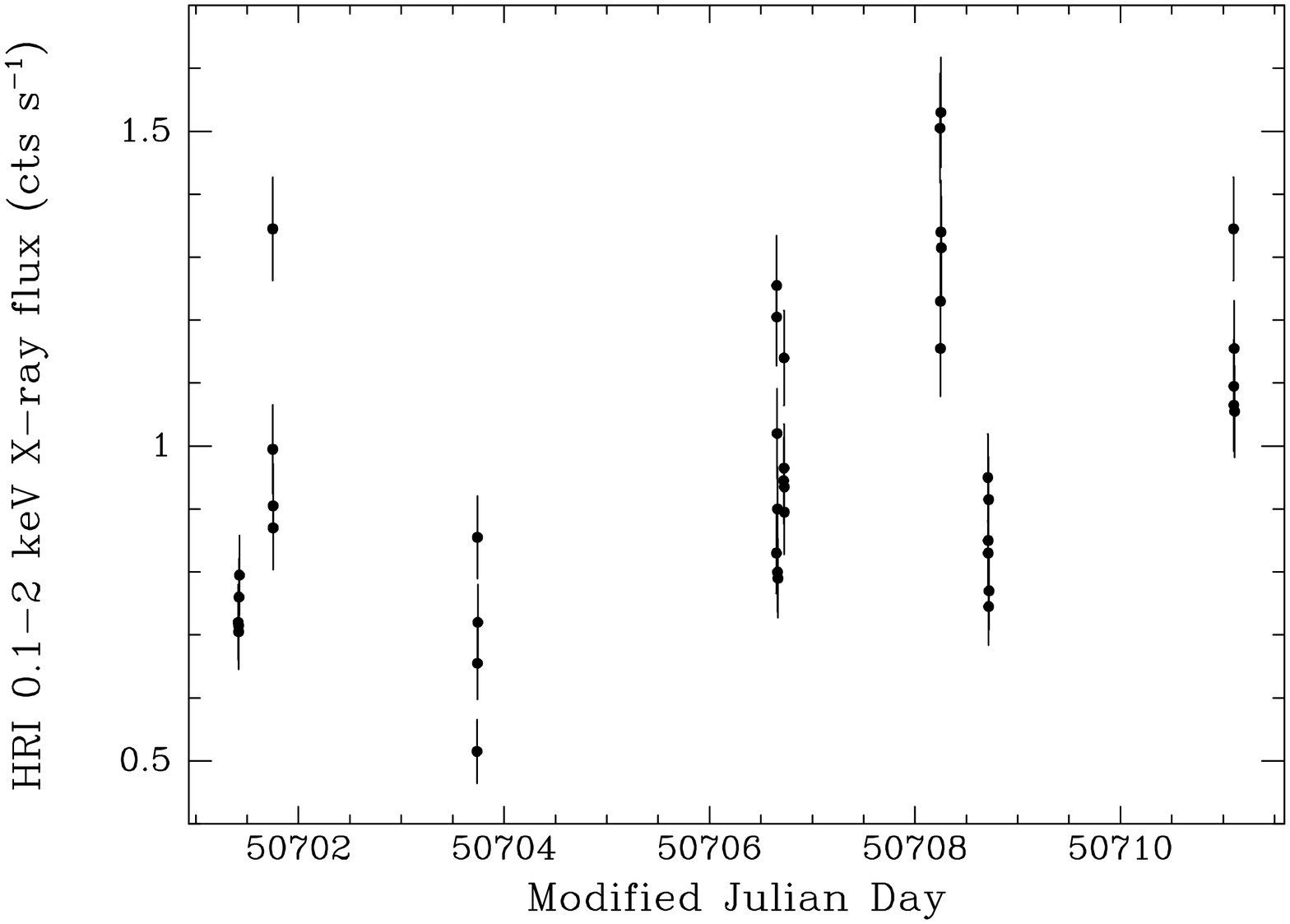,width=9.8cm,angle=-90}
\end{center}
\vspace{-3.8cm}
\caption[]{{\it (Upper panels)} 2--10 keV background-subtracted light
curves of {\it BeppoSAX} MECS (left) and {\it ASCA} GIS2+3 (right).
Both light curves are binned at 200 s.
{\it (Central panels)} 2--10 keV background-subtracted light curves
of {\it RXTE} PCA observations I (left) and II (right). 
Both light curves are binned at 16 s. The average uncertainties in the
flux measurements of the two pointings are indicated in the upper left and
upper right corners of left and right panels, respectively.
{\it (Lower panels)} 0.1--2 keV background-subtracted light curves of {\it
ROSAT} PSPC (left) and HRI (right) pointings. Both light
curves are binned at 200 s. Given that, during the PSPC observation,
good source events were only recorded in the first $\sim$2 hours and in
the last $\sim$5 hours of the pointing, we only plot these two time 
intervals in order to make the light curve of this observation clearer.
Fast (10--1000 s) variations are detected during all pointings.
0 hours UT of the day of the beginning of the observation (reported in
Table 1) were used as the reference time for panels in which times are
expressed in hours}
\end{figure*}

\subsection{X--ray data}

\subsubsection{Positional information}

Among all the pointed X--ray observations presented here, the most accurate
source position for 4U 1700+24 can be determined using the {\it
ROSAT} HRI data of the September 1997 pointing. 
The results of the HRI
observation on 4U 1700+24 show that the source is located at $\alpha$ =
17$^{\rm h}$ 06$^{\rm h}$ 34$\fs$49, $\delta$ = +23$^{\circ}$ 58$'$
18$\farcs$22 (J2000); the total (boresight plus statistical) error along 
both coodinates is 8$''$.

This position is in total agreement with the one provided in the
{\it ROSAT} PSPC bright source catalogue of Voges et al. (1999). 
This is true even if only the statistical uncertainty (0$\farcs$7) is
considered.
We also confirm, after comparison with the Digitized Sky Survey, that 
the only object brighter than the magnitude limit of this survey (i.e.
with $R \la$ 21.5) lying inside the HRI error circle is indeed the star 
V934 Her; again, even considering the statistical
error only, the HRI localization is fully consistent with the optical 
counterpart proposed by Garcia et al. (1983), the coordinates of which 
are given with an extreme degree of accuracy (the reported error is 
less than 0$\farcs$1 along both coordinates) in the {\it Hipparcos} 
catalogue (Perryman et al. 1997). This further strengthens the 
association between 4U 1700+24 and V934 Her.
This is in agreement, within the uncertainties, with the results
independently obtained by Morgan \& Garcia (2001) from this same {\it
ROSAT} HRI observation.

\subsubsection{Light curves and timing analysis}

The long-term (1995/2001) 2--10 keV variability of 4U 1700+24 is evident
from the {\it RXTE} ASM light curve reported in Fig. 1. The increase in
the 4U 1700+24 count rate detected in the Fall of 1997 is quite clear, as
well as long-term variations of the low state level. However, an apparent
recursive behaviour, with peaks which are in any case much less evident
than that of Fall 1997, seems to appear in the ASM light
curve after the aforementioned outburst. The timing analysis of the 5-day
averaged data shows a tentative periodicity of $\approx$400 days for the
outbursting activity of the source. A similar indication is obtained using
the dwell-by-dwell and the daily-averaged ASM data sets. Instead, the
suggested optical periodicity of 31.41 days reported by Perryman et al.
(1997) from {\it Hipparcos} observations is not found in any of the ASM 
X--ray data sets. 

We also examined the shorter-term X--ray light curves of the pointed
observations on 4U 1700+24. These are reported in Fig. 3. 
All curves are rebinned at 200 s with the exception
of those from the two {\it RXTE} pointings. These are binned with a
shorter time interval (16 s) as they allow detecting fast variations with
good S/N thanks to the PCA sensitivity.

As one can see, the variability timescale lies between $\sim$10 s
and several thousands of seconds.
The substantial erratic variations already detected with {\it EXOSAT}
(Dal Fiume et al. 1990) are clearly present in all observations. 
The source flux in the {\it BeppoSAX} pointing is significantly lower
than that in the {\it ASCA} and {\it RXTE} observations, and comparable
with that detected in the {\it EXOSAT} observation (see next Subsection).
Likewise, the {\it ROSAT} light curves at softer (0.1--2 keV) X--ray 
energies show an erratic variability behaviour similar to that present 
in the 2--10 keV light curves. Moreover, the longer HRI pointing, 
performed during the rising branch of the Fall 1997 outburst, shows a 
hint of increase in the emission with time.
These variations do not appear to be correlated with the source hardness,
this remaining fairly constant throughout each single pointing.

Timing analysis on the X--ray light curves from the various spacecraft
was performed with the FTOOLS\footnote{available at: \\ {\tt
http://heasarc.gsfc.nasa.gov/ftools/}} (Blackburn 1995)
task XRONOS, version 4.02, after having converted the event arrival
times to the solar system barycentric frame.
The erratic source variability is clearly visible in the Power
Spectral Density (PSD); however, no coherent pulsations or periodicities
are detected at any significant level in the $f \sim$ 10$^{-4}$--10 Hz
frequency range. 
No indication of the $\sim$900 s X--ray periodicity found by Morgan \&
Garcia (2001) from {\it Einstein} data is found in our data set.
Only a 1/$f$-type PSD distribution, typical of the shot-noise behaviour
evident from the light curves in Fig. 3, is found below 10$^{-2}$ Hz.
An example of this is shown in Fig. 4, where a PSD calculated on the
time series of the summed 2--10 keV GIS2 and GIS3 count rates binned at
0.1 s is reported. The power density spectra were calculated for runs with
typical duration of 3000 s. The PSD shown in Fig. 4 is obtained by
averaging the temporal spectra of different runs and by summing adjacent
frequencies with a logarithmic rebinning.

Likewise, a Fast Fourier Transform (FFT) analysis was made to search for 
fast periodicities (pulsations or fast QPOs) in the 1--2000 Hz range 
within the {\it RXTE} data of observations I and II.
In both observations, data in ``Good Xenon" mode with a 2$^{-20}$ s 
(i.e. 1 $\mu$s) time resolution and 256 energy bands were available.
To search for fast pulsations and/or kiloHertz QPOs we made several FFTs
of 16 s long data segments in the 2--10 keV and 10--60 keV energy
intervals and with a Nyquist frequency of 4096 Hz.
The power density spectra were then averaged for each observation.
To reduce the effect of any possible (albeit unlikely) short-term
orbital modulation we divided, for each observation, the whole data span
in subintervals $\sim$3000 s long (a typical {\it RXTE} contiguous data
segment). We calculated the $Z^2$ statistics (Buccheri et al. 1983)
on each subinterval and finally added together the results. No presence of
QPO peaks or of coherent pulsations was detected in both {\it RXTE} 
observations.

\begin{figure}
\begin{center}
\epsfig{figure=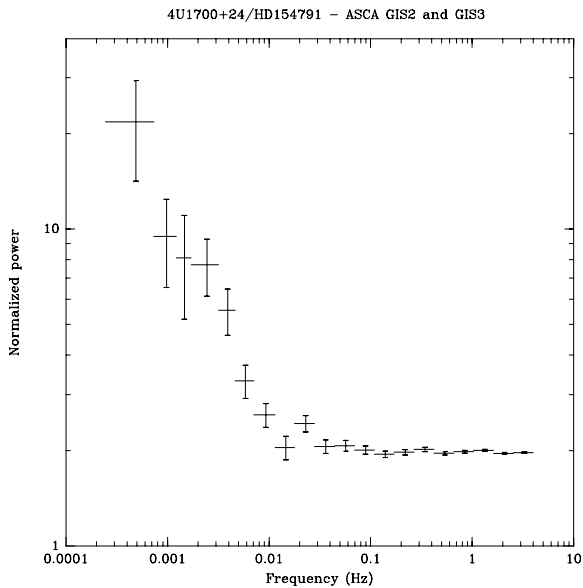,width=9cm}
\end{center}
\vspace{-1cm}
\caption[]{2--10 keV PSD of 4U 1700+24 obtained from the summed GIS2 and
GIS3 events from the March 1995 {\it ASCA} observation. The PSD is
normalized according to the prescription by Leahy et al. (1983).
No coherent oscillation or periodicity is found, while a shot-noise
1/$f$-type trend is apparent at low frequencies}
\end{figure}

\begin{table*}
\caption[]{Best-fit parameters for the X--ray spectra of 4U 1700+24 coming
from the observations described in this paper. In all models the 
hydrogen column density was fixed at the Galactic extinction value
$N_{\rm H}$ = 4$\times$10$^{20}$ cm$^{-2}$. Luminosities, corrected for
interstellar Galactic absorption, are computed assuming a distance $d$ =
420 pc and are expressed in units of 10$^{32}$ erg s$^{-1}$. In the cases
in which the X--ray data could not completely cover the X--ray interval of
interest for the luminosity determination, an extrapolation of the
best-fit model is applied. Errors and upper limits are at a 90\% confidence 
level for a single parameter of interest. Observations are reported in 
chronological order from left to right}
\begin{center}
\begin{tabular}{c|c|c|c|c|c|c}
\hline
\noalign{\smallskip}
Model & {\it EXOSAT} & {\it ROSAT} & {\it ASCA} & {\it RXTE} I & {\it
RXTE} II & {\it BeppoSAX} \\
parameter & (1--10 keV) & (0.1--2 keV) & (0.6--10 keV) & (3--20 keV) &
(3--100 keV) & (0.1--30 keV) \\
\noalign{\smallskip}
\hline
\noalign{\smallskip}
${\chi^{2}}$/dof & 21.9/29 & 23.0/16 & 258/246 & 26.5/41 & 71.6/98 &
68.9/64 \\
\multicolumn{1}{l|}{BB:} & & & & &\\
$kT_{\rm BB}$ (keV) & --- & 1.3$^{+1.5}_{-0.5}$ & 0.86$^{+0.07}_{-0.09}$&
1.24$^{+0.06}_{-0.04}$ & 1.0$^{+0.1}_{-0.5}$ & --- \\
$R_{\rm BB}$ (m) & --- & 50$^{+90}_{-30}$ & 65$^{+14}_{-8}$ & 
54$^{+14}_{-12}$ & 170$^{+100}_{-50}$ & --- \\
\multicolumn{1}{l|}{Comptonization:} & & & & & & \\
$kT_0$ (keV) & --- & --- & --- & --- & 1.4$^{+0.4}_{-0.3}$ & --- \\
$kT_{e^{-}}$ (keV) & 1.16$\pm$0.18 & 1.8($<$8.2) & 2.1$^{+0.7}_{-0.4}$ &
4.9$^{+1.7}_{-0.7}$ & 10.8$^{+3}_{-1.6}$ & 1.16$^{+0.05}_{-0.06}$ \\
$\tau$ & 32$^{+30}_{-7}$ & 9.5$^{+4.4}_{-0.6}$ & 25$^{+7}_{-6}$ & 11$\pm$2
& 2.5$\pm$0.4 & 34$^{+5}_{-2}$ \\
$Normalization^*$ & 2.3$^{+0.8}_{-1.1}$ & 1.0$^{+0.2}_{-0.3}$ &
1.41$^{+0.14}_{-0.12}$ & 10$\pm$3 & 7.4$^{+2.5}_{-2.4}$ &
1.52$^{+0.15}_{-0.16}$ \\
\noalign{\smallskip}
\hline
\noalign{\smallskip}
\multicolumn{1}{l|}{X--ray luminosity:} & & & & & & \\
0.1--2   keV  & ---  & 3.56 & ---  & ---  & ---  & 1.06  \\
  2--5   keV  & 1.89 & ---  & 3.17 & 8.89 & 43.9 & 1.40  \\
  5--10  keV  & 0.83 & ---  & 2.27 & 7.85 & 57.6 & 0.68  \\
  2--10  keV  & 2.72 & ---  & 5.45 & 16.7 & 101  & 2.07  \\
 10--100 keV  & ---  & ---  & ---  & ---  & 112  & 0.047 \\
\noalign{\smallskip}
\hline
\noalign{\smallskip}
\multicolumn{1}{l|}{BB/Comp. flux ratio:} & & & & & & \\
2--10 keV  & $<$0.05 & --- & 0.70 & 0.82 & 0.30 & $<$0.04 \\
0.1--2 keV & ---     & --- & 0.27 & ---  & ---  & $<$0.03 \\
\noalign{\smallskip}
\hline
\noalign{\smallskip}
\multicolumn{7}{l}{$^*$ in units of 10$^{-3}$; all normalizations
are relative to the {\sc compST} model with the exception of that of the
{\it RXTE}} \\
\multicolumn{7}{l}{observation II, which is instead referring to the 
{\sc compTT} model (see text).} \\
\noalign{\smallskip}
\hline
\end{tabular}
\end{center}
\end{table*}

\subsubsection{Spectra}

In order to perform spectral analysis, the pulse-height spectra from the
detectors of all spacecraft were rebinned to oversample by a factor 3
the full width at half maximum (FWHM) of the energy resolution and to
have a minimum of 20 counts per bin, such that the $\chi^2$ statistics
could reliably be used.
For all detectors, data were then selected, when a sufficient number of 
counts were obtained, in the energy ranges where the instrument responses 
are well determined. 
We then used the package {\sc xspec} (Arnaud 1996) v11.0 to fit the 
resulting broad band energy spectra.

In the broad band {\it BeppoSAX} fits, normalization factors were applied 
to LECS, HPGSPC and PDS spectra following the cross-calibration tests between 
these instruments and the MECS (Fiore et al. 1999). These factors were
constrained to be within the allowed ranges during the spectral fitting.
Similarly, a constant was introduced between GIS and SIS spectra to allow
intercalibration between the two GIS and the two SIS.
For the same reason we allowed for a calibration constant between PCA and 
HEXTE in the {\it RXTE} data spectral fits. Also, in order to keep into 
account systematic errors in the response matrices of PCA and HEXTE 
and the intercalibration uncertainties between these two detectors, a 
systematic uncertainty of 1\% in the count rate was added to their 
spectra before fitting. Moreover, the spectra from the two HEXTE clusters 
were fitted independently and the normalization between them was left free 
given the presence of a small systematic difference between their 
response.
A photoelectric absorption column, modeled using the Wisconsin
cross sections as implemented in {\sc xspec} (Morrison \& McCammon 1983)
and with solar abundances as given by Anders \& Grevesse (1989),
was applied to all spectral models used in the data analysis and
illustrated in this Subsection.

{\it BeppoSAX} and {\it RXTE} data show continuum emission above 10 keV
coming from the source; during outburst, {\it RXTE} detects the object up
to 100 keV. Moreover, on the low-energy end of the X--ray spectral range,
a clear soft excess below 1 keV is apparent from the {\it BeppoSAX} and
{\it ROSAT} observations (see Fig. 5). In order to find a suitable model
which could fit spectra of all observations we started with simple models
such as blackbody (BB), disk-blackbody (DBB; Mitsuda et al. 1984), power
law, cut-off power law and thermal bremsstrahlung.
None of these models could produce an acceptable fit to our X--ray data
with the only exception of the {\it EXOSAT} spectrum which, as already
illustrated by Dal Fiume et al. (1990), can be fit either with a power
law or with a thermal bremsstrahlung. 

Motivated by the suggested similarity in the X--ray temporal and
spectral behaviours between $\gamma$ Cas (Frontera et al. 1987; Haberl
1995) and 4U 1700+24, we next tried to fit the data with the {\sc xspec}
models {\sc raymond}, {\sc mekal} and {\sc vmekal} representing
emission from hot diffuse gas (Raymond \& Smith 1977; Mewe et al. 1985).
This choice was suggested from the work by Kubo et al. (1998) and Owens
et al. (1999), who respectively modeled {\it ASCA} and {\it BeppoSAX}
observations of $\gamma$ Cas with such a prescription.
Our application of these models to our data set is instead far from
satisfactory: we get $\chi^{2}_{\nu}$ values between 4 and 8 for all
observations even allowing for unplausibly low ($<$10$^{-11}$) metal
abundances with respect to the solar value. Therefore, we reject this
representation of the source X--ray spectrum also.
Moreover, the partially absorbed bremsstrahlung model used in a preliminary 
analysis of our {\it ASCA} and {\it BeppoSAX} spectra (Dal Fiume et al.
2000) gives unacceptable fits when applied to the other observations.

Thus, we tried a composition of two models. The best results are obtained
considering a BB plus a Comptonization model. 
For the latter, we used the first-order modelization by Sunyaev and \&
Titarchuk (1980; {\sc compST} in {\sc xspec}) for the low-state
observations (i.e. all but the {\it RXTE} II), and the more complete one
by Titarchuk (1994; {\sc compTT} in {\sc xspec}) for the outburst
pointing. This choice is basically due to two reasons: (i) the lower S/N
of the low state spectra does not allow a good constraining of some (or
all) parameters of the {\sc compTT} model; (ii) the {\it RXTE} II
observation detects the source up to 100 keV, thus enabling a better
sampling of the high-energy component and then a better determination of
all {\sc compTT} parameters. It should be noted that the fit to the {\it
RXTE} observation II data is not sensitive to the geometry of the
Comptonizing cloud: we therefore decided to assume the (default) spherical
geometry.

The BB+Comptonization description of the data allows acceptable fits of
all the spectra of our data set.
Best-fit parameters are reported in Table 3. The X--ray spectra with the
best-fit model superimposed to the data and the fit residuals are shown in
Fig. 5.
We remark that the absorption column density $N_{\rm H}$ could not be
reliably determined from any of the spectra of our set; therefore, we
fixed it to the Galactic value in the direction of 4U 1700+24, which is 
$N_{\rm H}$ = 4$\times$10$^{20}$ cm$^{-2}$ (Dickey \& Lockman 1990). 
This value is in agreement with the $N_{\rm H}$ determined from the
optical interstellar absorption along that direction (Schlegel et al.
1998) by using the empirical relationship by Predehl \& Schmitt (1995).

\begin{figure*}
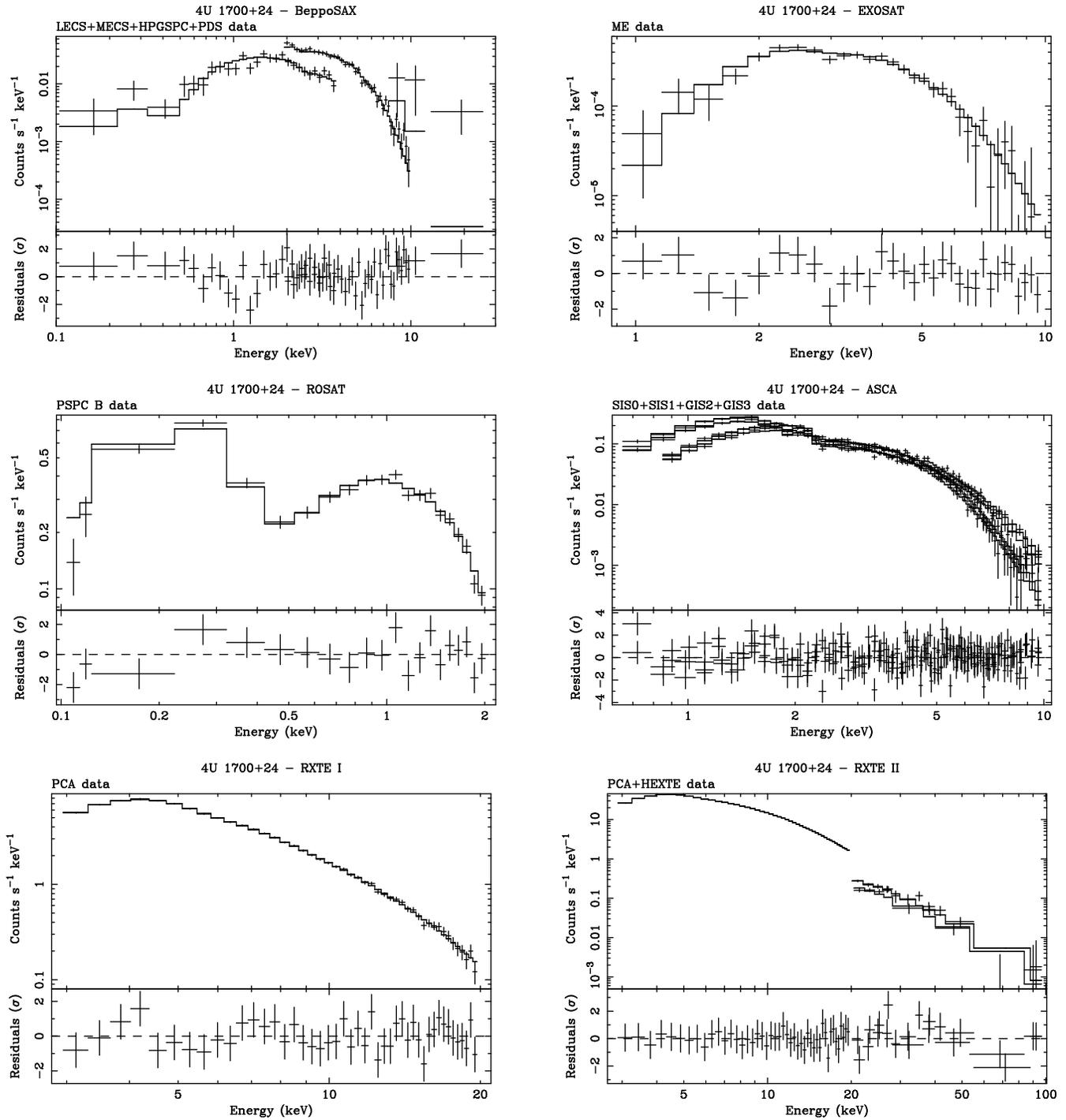

\begin{center}
\epsfig{figure=h3167f5ul.ps,width=6cm,angle=-90}
\hspace{1cm}
\epsfig{figure=h3167f5ur.ps,width=6cm,angle=-90}
\end{center}
\begin{center}
\epsfig{figure=h3167f5cl.ps,width=6cm,angle=-90}
\hspace{1cm}
\epsfig{figure=h3167f5cr.ps,width=6cm,angle=-90}
\end{center}
\begin{center}
\epsfig{figure=h3167f5ll.ps,width=6cm,angle=-90}
\hspace{1cm}
\epsfig{figure=h3167f5lr.ps,width=6cm,angle=-90}
\end{center}
\caption[]{X--ray spectra of 4U 1700+24 obtained from the spacecraft
pointed observations described in the text. The fit residuals using the
best-fit models reported in Table 3 are also presented for each
observation. In detail, the spectra in the upper row ({\it BeppoSAX},
0.1--30 keV, on the left, and {\it EXOSAT}, 1--10 keV, on the right)
are fit with a single, photoelectrically absorbed, Comptonization model;
all the other ones in the central ({\it ROSAT}, 0.1--2 keV, left, and {\it
ASCA}, 0.6--10 keV, right) and lower ({\it RXTE} I, 3--20 keV, left, and
{\it RXTE} II, 3--100 keV, right) rows are fitted with a photoelectrically
absorbed BB+Comptonization model}
\end{figure*}

As it can be seen from Table 3, the {\it BeppoSAX} and {\it EXOSAT} spectra
(which were taken at the lowest flux levels observed from the source 
in the present data set) the BB component is not required; for 
these two spectra, the fit is performed using a {\sc compST} model
only, and a 90\% confidence level upper limit to the BB flux was computed
(see Table 3). In all other cases the BB component is always detected, and 
the fits performed with this model plus a Comptonization are always much 
better than those obtained with either single Comptonization or 
DBB plus Comptonization models.

From our fits there appears to be no need for including an iron emission
line around 6.7 keV. In each observation we however computed the 90\% 
confidence level upper
limits for the equivalent width (EW) of a Fe emission at 6.7 keV assuming
different line FWHM. The results are reported in Table 4. As one can see,
the most stringent limits (coming from {\it RXTE} II and also, in the
narrow-line hypothesis, from {\it ASCA} observations) imply an EW less
than 25 eV in the case of a narrow Fe emission and less than 50 eV
considering a broad emission line.

\begin{table}
\caption[]{List of upper limits to the EW of the 6.7 keV Fe emission line,
assuming different fixed line FWHMs, for the various X--ray observations
covering that spectral energy. Pointings are listed in chronological
order from top to bottom}
\begin{center}
\begin{tabular}{lrrr}
\noalign{\smallskip}
\hline
\noalign{\smallskip}
\multicolumn{1}{c}{Observation} & \multicolumn{3}{c}{FWHM} \\
\noalign{\smallskip}
\cline{2-4}
\noalign{\smallskip}
 & \multicolumn{1}{c}{0.1 keV} & \multicolumn{1}{c}{0.5 keV} &
\multicolumn{1}{c}{1 keV} \\
\noalign{\smallskip}
\hline
\noalign{\smallskip}

{\it EXOSAT}   & $<$600  & $<$800  & $<$1380 \\
{\it ASCA}     & $<$24.4 & $<$53.9 & $<$163  \\
{\it RXTE} I   & $<$53.5 & $<$82.0 & $<$208  \\
{\it RXTE} II  & $<$24.4 & $<$31.6 & $<$49.7 \\
{\it BeppoSAX} & $<$170  & $<$470  & $<$1290 \\

\noalign{\smallskip}
\hline
\noalign{\smallskip}
\end{tabular}
\end{center}
\end{table}

\section{Discussion}

\subsection{X--ray variability: accretion onto a NS}

The X--ray light curves of 4U 1700+24 illustrated in Figs. 1 and 3, along
with the results shown in Table 3 for the various spectral observations
presented in this paper, confirm that this source has substantial X--ray
variability. In particular, we note that, as the X--ray luminosity
$L_{\rm X}$ (say, in the 2--10 keV band) increases, the emission from the
object becomes harder: the (5--10 keV)/(2--5 keV) hardness ratio spans
from $\sim$0.5 at the lowest flux level sampled here (the {\it BeppoSAX}
observation) to $\sim$1.3 during outburst (the {\it RXTE} II pointing);
likewise, the 10--100 keV flux rises of a factor more than 2000 between
these two observations, becoming stronger than the 2--10 keV flux during
outburst.
The fit parameters, in particular the temperature $kT_{\rm BB}$ and radius 
$R_{\rm BB}$ of the BB component, as well as the temperature $kT_{e^{-}}$ 
and the optical depth $\tau$ of the Comptonizing cloud, also show clear
correlations (direct for the first three parameters, inverse for the 
latter one) with $L_{\rm X}$ (see Table 3).

The ratio between the BB and Comptonization components has instead a 
more complex behaviour: it increases from basically zero to about 0.7
following the X--ray luminosity level in the low-intensity state, and
then drops to 0.3 during outburst.

All this is consistent with a model in which enhancement of accreting
matter onto a compact object is responsible for the observed spectral
shape and variability. Indeed, the two components detected in the X--ray
spectrum are suggestive of the presence of a compact object (responsible
for the BB emission) which is accreting matter and which is surrounded
by a corona of hot electrons; the latter emits hard X--rays by
Comptonization of the softer BB photons.
In addition, the hardening of the spectrum with increasing
luminosity seems to point toward accretion of matter with low specific
angular momentum, as suggested by e.g. Smith et al. (2001) and Wu et al.
(2002) to explain a similar behaviour observed in other X--ray binaries.

On the contrary, short-period hardness ratio variations are not detected
within each single pointed observation. This fast and ``colorless"
variability is typical of inhomogeneities inside the accretion flow (e.g.
van der Klis 1995). 
This explanation for the fast X--ray variability of 4U 1700+24 is also 
suggested by the ``shot-noise" 1/$f$-type PSD associated with all the 
pointed observations presented here.

Both these longer and shorter term variabilities would point to the
presence of variable accretion in this system.
The X--ray flux is also much harder and stronger than that expected from
coronal emission of a late-type giant star, this being generally 2--3
orders of magnitude lower (H\"unsch et al. 1998) than the one detected
here even at the lowest luminosity level.

Thus, the X--ray spectral shape and the temperatures of the BB and the
Compton cloud would suggest the presence of an extremely compact accreting
object, most likely a NS or a BH.
Indeed, a WD as accretor is basically ruled out because of the X--ray
spectral shape, which (as stated in the previous Section) is completely
different from that observed by Kubo et al. (1998) and Owens et al. (1999)
in the system $\gamma$ Cas, which is thought to host an accreting WD.
Moreover, the high BB temperatures
found here are incompatible with those of a WD surface, which basically
emits in the UV rather than in the soft X--rays. Also, the X--ray spectrum
of 4U 1700+24 observed in outburst is definitely too hard to be emitted by
an accreting WD. The BH interpretation for the nature of the accreting
compact object also seems less strong than the NS solution because 
the hard component of the X--ray spectra reported here
is generally systematically softer than what expected in the case
of an accreting BH (e.g. Barret et al. 2000).
So, we suggest that the accreting object is a NS, even if no pulsations or
X--ray bursts were ever detected from this source.

It is evident from Table 3 that in every observation the size of the
emitting area of the BB is remarkably small, of the order of tens of
metres. Shortcomings and limitations of the BB model are known
(Gierli\'nski et al. 1999; Merloni et al. 2000); however, these would
hardly compensate for the entire difference between the inferred
($\sim$200 m) BB emitting area and that expected in the case of isotropic
BB emission from a NS (few km). We therefore suggest that accretion in
this system does not take place in a spherical fashion onto the compact
object but, rather, that mass coming from
the M-type companion in the form of a stellar wind is captured by the NS
and then accreted onto a small zone of its surface. An equatorial
accretion belt (such as a boundary layer) is disfavoured because the wind
has no high intrinsic angular momentum, so a disk can hardly be formed
with this accretion mechanism. Indeed, the use of a DBB model instead
of a BB produces worse (and in some cases unacceptable) fits to our X--ray
data.
Also, the non-detection kiloHertz QPOs (which are believed to originate in
the inner parts of the disk; see e.g. van der Klis 2000) and of enhanced
optical activity (e.g. emission lines) during the Fall
1997 outburst points to the absence of an accretion disk as usually
observed in LMXRB where accretion onto the compact object takes place via
Roche-lobe overflow of the secondary.
Besides, as already remarked by Garcia et al. (1983), in this system the
matter transfer onto the compact object is hardly occurring via Roche-lobe
overflow as this would produce X--ray emission which is 3--4 orders of
magnitude more intense than observed.

An alternative explanation comes by considering a magnetically-driven 
accretion scenario. In this case, if the magnetic field of the compact 
object is strong enough, two accretion columns would form and the captured 
matter flows along field lines and impacts the NS surface in correspondence 
of the magnetic polar caps.
This would however produce pulsations if the rotational and magnetic
field axes are not aligned. Given that we do not detect X--ray pulsations
from 4U 1700+24, the alignment between these two axes must be invoked to
explain the observations. Although this would appear a rather 
{\it ad hoc} hypothesis, we must remark that a similar interpretation has 
been proposed to explain the absence of pulsations in 4U 1700$-$37 (White
et al. 1983) and, more recently, in 4U 2206+54 (Corbet \& Peele 2001), which
are believed to host an accreting NS.

Then, if the NS rotation axis is coincident with the magnetic field axis,
and the BB emitting area is that of the polar caps only, we could easily
obtain, for the emitting area around the NS poles, a size of the order of 
the BB radius measured from the X--ray spectral fits.
Alternatively, we can explain the lack of pulsations from the magnetized
accreting NS by assuming a low inclination angle for the system (see
next Subsection) and the NS rotational axis perpendicular to the orbital
plane of the system. In this case, even assuming a non-zero (say,
$\approx$20$^{\circ}$) angle between the NS magnetic field and rotational
axes, the system geometry is such that we continuously see X--ray
radiation from a single polar cap of the NS. Clearly, the emission will
not be modulated by the NS rotation in this case also.

We must however note that a further possible explanation for the small BB
emitting area size is the following. The estimate coming from our fits is
based on the assumption of isotropic emission from the NS surface. It has
however been shown (London et al. 1986) that because of cooling and
back-warming effects the spectrum at the NS surface, if fitted with a 
``classic" BB model, 
can lead to the net effect of overestimating the BB emitting area by
as much as 2 orders of magnitude.
This would possibly explain our fit results without invoking the funneling
of matter onto the NS polar caps by means of an intense magnetic field and
the alignment of its axis with the NS rotation axis. In this case, 
one can overcome the very low ($\sim$10$^{-5}$ -- 10$^{-6}$)
accretion efficiency assuming that the NS is orbiting the M giant in a 
wide orbit (i.e. with separation $a \ga$ 500 $R_\odot$; see below), and 
thus only a very small quantity of the emitted stellar wind can be 
captured and accreted by the compact object. 

\subsection{A possible scenario for the system}

In the hypothesis that the observed X--ray luminosity coming from 4U
1700+24 is due to accretion, an inferred accretion rate onto the NS of
$\dot{M} \approx$ 10$^{-14}$ $M_\odot$ yr$^{-1}$ is found.
This can be explained by assuming a typical mass loss rate via
stellar wind of a red giant ($\dot{M} \approx$ 10$^{-9}$ $M_\odot$
yr$^{-1}$; e.g. Willson 2000) and an accretion efficiency $\eta$ =
10$^{-4}$ which is quite normal for a NS accreting from the
stellar wind (e.g. Frank et al. 1992). The remaining difference between 
the two above values of $\dot{M}$ can be accounted for by a partial
inhibition of the accretion due to the ``propeller effect" 
(Illarionov \& Sunyaev 1975), according to which the magnetosphere
of the NS acts as a barrier to accretion of matter onto the NS
surface. Nevertheless, a small fraction of matter can flow along the
magnetic field lines and find a way to the NS. This ``leaking" of
matter through the magnetosphere has been invoked to explain the
X--ray emission of A0535+26 during its low intensity state (Negueruela et
al. 2000) and, more recently, of 4U 2206+54 (Corbet \& Peele 2001).

The presence of a cold wind from the secondary can be suggested by 
three further issues: the UV variability, the possible additional
absorption detected by Gaudenzi \& Polcaro (1999) in the bluest part of
their optical spectra, and the fact that the {\it IRAS} measurements of
the mid-infrared source flux at 12 and 25 $\mu$m are about a factor three
larger than those expected from a normal M giant.
Modulations in the wind density, such as density waves produced by pulsations 
of the M giant envelope, could be responsible for the tentatively observed 
periodicity in the ASM X--ray light curve.
Alternatively, the compact object and the M giant are orbiting around each 
other along a wide and eccentric orbit, and the X--ray enhancements are due to
the transit of the NS through the inner and denser parts of the stellar
wind. It is not however clear why no indication of X--ray emission
enhancements were present before the large X--ray outburst of Fall 1997.
Possibly, this event triggered the mechanism of producing these
``periodic" increases of X--ray activity even though at a lower level with
respect to the first one.

Whichever the reason for these phenomena, accretion from higher density
regions of the stellar wind onto an extremely compact object (very likely
a NS) is the most viable explanation for the observed behaviour of
luminosity and spectral parameters in the X--ray range. These density
variations would modulate the strength of the propeller effect;
indeed, the magnetospheric radius is expected to be larger (and thus the
magnetospheric barrier should be stronger) when the density of the
accreting matter is lower, thus better inhibiting accretion onto the NS. 
The faster (10--1000 s) variations in the X--ray luminosity are instead most 
probably produced by small-scale inhomogeneities of the accretion flow
onto the compact object.

Concerning the ``normal" spectrum of the optical counterpart, 
V934 Her, we note that the bolometric luminosity (most of which is emitted
in the optical and near-infrared bands) of a M2 III giant is 550
$L_\odot$ (Lang 1992), which is a factor $\sim$200 higher than the
$L_{\rm X}$ measured during outburst from 4U 1700+24 (see Table 3).
This would explain the absence in the optical spectrum of emission features 
(Tomasella et al. 1997) which could be induced by the possible X--ray 
irradiation during outburst. For instance, the object GX 1+4, which is
often compared to 4U 1700+24 as it is the only other LMXRB with a M giant
secondary, displays a composite optical spectrum with numerous emission
lines (Chakrabarty \& Roche 1997); it however has a bolometric flux
from the secondary which is 4 times less than the X--ray luminosity of
the compact object (a pulsating NS). This can naturally explain the
difference between the optical spectra of these two X--ray binaries.
Possibly, the emission wings from Balmer and He lines detected by Gaudenzi
\& Polcaro (1999) in V934 Her could be an indication of emission coming
from small regions of the accretion flow or of the M giant surface heated
by the X--ray irradiation.

Next, we would like to briefly discuss about the possible size and orbital 
period of this system. In order to match the very low ($<$3 km s$^{-1}$)
projected orbital velocity of the secondary star obtained from optical
spectroscopic measurements (Garcia et al. 1983) with the fact that the
accretion most likely occurs via stellar wind capture, the system should
be fairly wide. Indeed, considering $P_{\rm orb} \approx$ 400 days and
assuming typical masses for a NS (1.4 $M_\odot$) and a M2 III star (1.3
$M_\odot$; Lang 1992), we get by applying Kepler's third law an orbital
separation $a \approx$ 300 $R_\odot$, thus substantially larger than the
radius of a M2 III giant ($\approx$ 60--70 $R_\odot$; Dumm \& Schild
1998).
In this case the orbital velocity would however be $\sim$10 times larger
than the upper limit to its projected value measured by Garcia et
al. (1983).
Thus, the above prescription would be valid only if the system has a low
inclination angle ($i <$ 10$^{\circ}$).
Alternatively, considering higher vales for $i$ and keeping the orbital
separation larger than the radius of a M2 III star, we obtain longer
(2--5 years) orbital periods. In the latter case, the tentative 400-day
X--ray activity periodicity of the source would be tied to density waves
in the stellar wind and thus to possible secondary star pulsations, rather
than to the orbital modulation.

Finally, we want to draw the attention of the reader on the fact
that this low-luminosity source can be studied in the high energy
range because of its proximity. Indeed, if it were at, say, 4 kpc from
earth, its emission would have been hardly detected by any of
the spacecraft quoted above. Therefore, several other systems similar to
4U1700+24/V934 Her might be present throughout the Galaxy, and its
``uniqueness" is actually a selection effect due to a combination of its
low X--ray emission and its closeness.

\section{Conclusions}

We analyzed the existing and mostly still unpublished data set of X--ray
pointed observations on the source 4U 1700+24. These observations span a
time baseline of more than 15 years and were acquired with 5 different
spacecraft. In parallel, we presented (quasi-)simultaneous ground-based
optical spectroscopic observations. 

The optical data allow us to show that the optical counterpart is a
red giant of spectral type M2 III with no significant peculiarities.
With this more accurate classification we can determine a better
distance estimate to the object, which lies at $\sim$400 pc from earth.
The X--ray data show strong and fast erratic shot-noise variability with
no hardness variations connected with the intensity changes. 
On the longer term, instead, the X--ray emission from the source becomes
harder as the luminosity increases. No pulsations, periodicities or
quasi-periodic oscillations (QPOs) are detected in the X--ray light curve
with the possible exception of a tentative long-term periodicity of 
$\approx$400 days. 
X--ray spectra can generally be well described using a BB+Comptonization
model, although the BB component is practically undetected at the lowest
X--ray intensity states. This spectral modelization is suggestive of a
compact object harboured in this system and accreting matter from the
giant M star positionally coincident with the X--ray localization. The
rather high BB and Comptonizing cloud temperatures along with the X--ray
spectral shape of the Comptonization component suggest the presence of a
NS as the accreting collapsed object.

The observed 2--10 keV X--ray luminosity of the source ranges from
$\sim$2$\times$10$^{32}$ during the low intensity state and
$\sim$1$\times$10$^{34}$ erg s$^{-1}$ at the peak of the outburst 
which occurred
in the Fall of 1997. These values point to accretion from a stellar wind
rather than via Roche-lobe overflow, in which case one would expect X--ray
luminosities larger by 3--4 orders of magnitude. The fast X--ray
variations are in this hypothesis connected with inhomogeneities in the
accreted wind.

The markedly small size of the BB emitting area inferred from our fits to
the X--ray data could indicate that wind accretion is funneled by the NS
magnetic field onto the magnetic polar caps; given that no X--ray 
pulsations are detected from the object, it would be possible that the
magnetic and rotational axes of the NS are coincident, or that the system 
has a very low inclination. In addition, the magnetic propeller effect 
would explain the very low accretion efficiency.
Alternatively, cooling effects on the NS surface could explain this small
BB emitting area, without invoking a substantially strong magnetic field 
for the collapsed object; in this case, the low accretion efficiency 
may be interpreted by assuming a wide binary separation for the system.
Indeed, given the known upper limit to any orbital motion of the M giant,
only systems with wide (hundreds of $R_\odot$) separations could be fit
into the outcoming observational picture. 

\begin{acknowledgements}
This work is supported by the Agenzia Spaziale Italiana (ASI) and the
Italian Consiglio Nazionale delle Ricerche (CNR). {\it BeppoSAX} is a
joint program of ASI and of the Netherlands Agency for Aerospace Programs
(NIVR). The {\it ASCA} observation was performed as part of the joint
ESA/Japan scientific program.
This research has made wide use of data obtained through the High Energy
Astrophysics Science Archive Research Center Online Service, provided by
the NASA/Goddard Space Flight Center.
This research has also made use of the SIMBAD database, operated at CDS,
Strasbourg, France. ASM data were provided by the {\it RXTE} ASM teams at
MIT and at the {\it RXTE} SOF and GOF at NASA's GSFC. We thank the 
anonymous referee for valuable comments. We also thank Massimo Cappi 
and Mauro Dadina for assistance in the {\it ASCA} data analysis, Stefano 
Bernabei for useful discussions on the optical spectra, and the staff 
astronomers of the Observatory of Bologna in Loiano for assistance 
during observations.
\end{acknowledgements}


\begin{thebibliography}{}

\bibitem{} Anders, E., \& Grevesse, N., 1989, Geochim. Cosmochim. Acta,
	53, 197

\bibitem{} Arnaud, K.A., 1996, {\sc xspec}: the first ten years. In: 
	Jacoby, G.H., \& Barnes, J., (eds.) ``Proceedings of the V ADASS
	Symposium", ASP Conf. Ser., 101, 17

\bibitem{} Barret, D., Olive, J.F., Boirin, L., et al., 2000, ApJ, 533,
	329

\bibitem{} Blackburn, J.K., 1995. In: Shaw, R.A., Payne, H.E., Hayes,
	J.J.E., (eds.) Astronomical Data Analysis Software and Systems
	IV, ASP Conf. Ser., 77, 367 

\bibitem{} Boella, G., Butler, R.C., Perola, G.C., et al., 1997a, A\&AS,
	122, 299

\bibitem{} Boella, G., Chiappetti, L., Conti, G., et al., 1997b, A\&AS,
	122, 327

\bibitem{} Bradt, H.V., Rothschild, R.E., \& Swank, J.H., 1993, A\&AS, 97,
	355

\bibitem{} Buccheri, R., Bennett, K., Bignami, G.F., et al., 1983, A\&A, 
	128, 245

\bibitem{} Chakrabarty, D., \& Roche, P., 1997, ApJ, 489, 254

\bibitem{} Chiappetti, L., \& Dal Fiume, D., 1997, The XAS Data Analysis
	System. In: di Ges\`u, V., Duff, M.J.B., Heck, A., Maccarone, 
	M.C., Scarsi, L., \& Zimmermann, H.-U., (eds.) ``Proceedings of
	the Fifth International Workshop on Data Analysis in Astronomy",
	World Scientific Press, 101

\bibitem{} Cooke, B.A., Ricketts, M.J., Maccacaro, T., et al., 1978,
	MNRAS, 182, 489

\bibitem{} Corbet, R.H.D., \& Peele, A.G., 2001, ApJ, in press 
	(astro-ph/0107131)

\bibitem{} Dal Fiume, D., Poulsen, J.M., Frontera, F., et al., 1990, Il
	Nuovo Cimento, 13 C, 481

\bibitem{} Dal Fiume, D., Masetti, N., Bartolini, C., et al., 2000, ASCA
	and BeppoSAX observations of the peculiar X--ray source 
	4U 1700+24/HD154791. In: McConnell, M.L., \& Ryan, J.M., (eds.) The
	Fifth Compton Symposium, AIP Conf. Proc., 510, 236 (astro-ph/0002348) 

\bibitem{} Dickey, J.M., \& Lockman, F.J., 1990, ARA\&A, 28, 215

\bibitem{} Diodato, M., 1998, ``Laurea" Degree Thesis, Dipartimento di 
	Astronomia, Universit\`a di Bologna

\bibitem{} Dumm, T., \& Schild, H., 1998, NewA, 3, 137

\bibitem{} Engvold, O., \& Rygh, B.O., 1978, A\&A, 70, 399

\bibitem{} Fiore, F., Guainazzi, M., \& Grandi, P., 1999, Technical Report
	1.2, {\it BeppoSAX} scientific data center, available online at: 
	{\tt ftp://www.sdc.asi.it/pub/sax/doc/\\
      software\_docs/saxabc\_v1.2.ps}

\bibitem{} Forman, W., Jones, C., Cominsky, L., et al., 1978, ApJS, 38,
	357

\bibitem{} Frank, J., King, A.R., \& Raine, D.J., 1992, Accretion Power in
	Astrophysics. Cambridge Univ. Press, Cambridge

\bibitem{} Frontera, F., Dal Fiume, D., Robba, N.R., et al., 1987, ApJ,
	320, L127

\bibitem{} Frontera, F., Costa, E., Dal Fiume, D., et al., 1997, A\&AS,
	122, 357

\bibitem{} Garcia, M.R., Baliunas, S.L., Doxsey, R., et al., 1983, ApJ,
	267, 291

\bibitem{} Gaudenzi, S., \& Polcaro, V.F., 1999, A\&A, 347, 473

\bibitem{} Gierli\'nski, M., Zdziarski A.A., Poutanen, J., et al., 1999,
	MNRAS, 309, 496

\bibitem{} Gunn, J.E., \& Stryker, L.L., 1983, ApJS, 52, 121

\bibitem{} Haberl, F., 1995, A\&A, 296, 685

\bibitem{} Horne, K., 1986, PASP, 98, 609

\bibitem{} H\"unsch, M., Schmitt, J.H.M.M., Schr\"oeder, K.-P.,
	\& Zickgraf, F.-J., 1998, A\&A, 330, 225

\bibitem{} Illarionov, A.F., \& Sunyaev, R.A., 1975, A\&A, 39, 185

\bibitem{} Jacoby, G.H., Hunter, D.A., \& Christian, C.A., 1984, ApJS, 56,
	257

\bibitem{} Jahoda, K., Swank, J.H., Stark, M.J., et al., 1996. In:
	Siegmund O.H.W., \& Gummin M.A., (eds.) EUV, X--ray and Gamma-ray
	Instrumentation for Space Astronomy VII, Proc. SPIE 2808, 59

\bibitem{} Kazarovets, E.V., Samus, N.N., Durlevich, O.V., et al., 1999,
	IBVS 4659

\bibitem{} Kubo, S., Murakami, T., Ishida, T., \& Corbet, R.H.D., 1998,
	PASJ, 50, 417

\bibitem{} Lang, K.R., 1992, Astrophysical Data: Planets and Stars.
	Springer-Verlag, New York

\bibitem{} Lammers, U., 1997, The SAX/LECS Data Analysis System User
        Manual, SAX/LEDA/0010

\bibitem{} Leahy, D.A., Darbro, W., Elsner, R.F., et al., 1983, ApJ, 266,
	160

\bibitem{} Levine, A.M., Bradt, H.V., Cui, W., et al., 1996, ApJ, 469, L33

\bibitem{} Liu, Q.Z., van Paradijs, J., van den Heuvel, E.P.J., 2001,
	A\&A, 368, 1021

\bibitem{} London R.A., Taam, R.E., \& Howard, W.M., 1986, ApJ, 306, 170

\bibitem{} Manzo, G., Giarrusso, S., Santangelo, A., et al., 1997, A\&AS,
	122, 341

\bibitem{} Makishima, K., Tashiro, M., Ebisawa, K., et al., 1996, PASJ, 
	48, 171

\bibitem{} Merloni, A., Fabian, A.C., \& Ross, R.R., 2000, MNRAS, 313, 193

\bibitem{} Mewe, R., Groenschild, E.H.B.M., \& van den Oord, G.H.J., 1985,
	A\&AS, 62, 197

\bibitem{} Mitsuda, K., Inoue, H., Koyama, K., et al., 1984, PASJ, 36, 741

\bibitem{} Morgan, Jr., W.A., \& Garcia, M.R., 2001, PASP, 113, 1386

\bibitem{} Morrison, R., \& McCammon, D., 1983, ApJ, 270, 119

\bibitem{} Negueruela, I., Reig, P., Finger, M.H., \& Roche, P., 2000,
	A\&A, 356, 1003

\bibitem{} Ohashi, T., Ebisawa, K., Fukazawa, Y., et al., 1996, PASJ, 48,
	157

\bibitem{} Owens, A., Oosterbroek, T., Parmar, A.N., et al., 1999, A\&A,
	348, 170

\bibitem{} Parmar, A.N., Martin, D.D.E., Bavdaz, M., et al., 1997, A\&AS,
	122, 309

\bibitem{} Perryman, M.A.C., Lindegren, L., Kovalevsky, J., et al., 1997,
	A\&A, 323, L49; ESA, The Hipparcos and Tycho Catalogues, ESA 
	SP-1200 

\bibitem{} Pfeffermann, E., \& Briel, U.G., 1986, Proc. SPIE, 597, 208

\bibitem{} Pfeffermann, E., Briel, U.G., Hippmann, H., et al., 1987,
	Proc. SPIE, 733, 519

\bibitem{} Philip, A.G.D., \& Hayes, D.S., 1983, ApJS, 53, 751

\bibitem{} Predehl, P., \& Schmitt, J.H.M.M., 1995, A\&A, 293, 889

\bibitem{} Raymond, J.C., \& Smith, B.W., 1977, ApJS, 35, 419

\bibitem{} Rothschild, R.E., Blanco, P.R., Gruber, D.E., et al., 1998,
	ApJ, 496, 538

\bibitem{} Schaefer, B.E., 1986, PASP, 98, 556

\bibitem{} Schlegel, D.J., Finkbeiner, D.P., \& Davis, M., 1998, ApJ, 500,
	525

\bibitem{} Serlemitsos, P.J., Jalota, L., Soong, Y., et al., 1995, PASJ,
	47, 105

\bibitem{} Silva, D.R., \& Cornell, M.E., 1992, ApJS, 81, 865

\bibitem{} Smith, D.M., Heindl, W.A., Swank, J.H., 2001, ApJ, in press
	(astro-ph/0103304)

\bibitem{} Sunyaev, R.A., \& Titarchuk, L., 1980, A\&A, 86, 121

\bibitem{} Tanaka, Y., Inoue, H., \& Holt S.S., 1994, PASJ, 46, L37

\bibitem{} Titarchuk, L., 1994, ApJ, 434, 570

\bibitem{} Tomasella, L., Munari, U., Tomov, T., et al., 1997, IBVS 4537

\bibitem{} Turner, M.J.L., Smith, A., \& Zimmermann, H.-U., 1981, Space
	Sci. Rev., 30, 513

\bibitem{} Turnshek, D.E., Turnshek, D.A., Craine, E.C., \& Boeshaar,
	P.C., 1985, An atlas of digital spectra of cool stars. Western
	Research Company, Tucson

\bibitem{} Tr\"umper, J., 1982, Adv. Space Res., 2(4), 241

\bibitem{} van der Klis, M., 1995, Rapid aperiodic variability in X--ray
	binaries. In: Lewin, W.H.G., van Paradijs, J., \& van den Heuvel,
	E.P.J., (eds.) ``X--ray Binaries", Cambridge Univ. Press, 
	Cambridge, p. 252

\bibitem{} van der Klis, M., 2000, ARA\&A, 38, 717

\bibitem{} Voges W., Aschenbach B., Boller T., et al., 1999, A\&A, 349,
	389

\bibitem{} Wendker, H.J., 1995, A\&AS, 109, 177

\bibitem{} White, N.E., Kallman, T.R., \& Swank, J.H., 1983, ApJ, 269, 264

\bibitem{} Willson, L.A., 2000, ARA\&A, 38, 573

\bibitem{} Wu, K., Soria, R., Campbell-Wilson, D., et al., 2002, ApJ, 564,
	in press (astro-ph/0109222)

\bibitem{} Yamashita, A., et al., 1997, IEEE Trans. Nucl. Sci., 44, 847

\end{thebibliography}
\end{document}